\documentclass[fleqn,usenatbib]{mnras}
\usepackage{newtxtext,newtxmath}
\usepackage[T1]{fontenc}
\usepackage{graphicx}
\usepackage{threeparttable}
\usepackage{natbib}
\usepackage{float}
\usepackage{color}
\usepackage{bm}
\usepackage{subfigure}
\usepackage[figuresright]{rotating}
\usepackage{multirow}
\usepackage{url}
\usepackage{orcidlink}
\usepackage{upgreek}
\DeclareRobustCommand{\VAN}[3]{#2}
\let\VANthebibliography\thebibliography
\def\thebibliography{\DeclareRobustCommand{\VAN}[3]{##3}\VANthebibliography}

\title[Hubble WFC3 Spectroscopy of LTT-9779~b]{HST/WFC3 Constraints on the Abundances of OH and FeH in the Atmosphere of the Ultra-Hot Neptune LTT-9779 b
} 

\author[Zhou et al.]{
Li Zhou\orcidlink{0000-0003-2391-0093},$^{1}$
Xinyue Ma\orcidlink{0009-0009-7398-2942},$^{2,3}$
Bo Ma\orcidlink{0000-0002-0378-2023},$^{2,3}$\thanks{E-mail:mabo8@mail.sysu.edu.cn}
Wei Wang\orcidlink{0000-0002-9702-4441},$^{4}$\thanks{E-mail:wangw@nao.cas.cn}
Chengzi Jiang\orcidlink{0000-0003-1381-5527},$^{5,6}$
Enric Pallé,$^{5,6}$
\newauthor
Yonghao Wang\orcidlink{0000-0003-0261-6362},$^{7}$
Jinpeng Wang,$^{8,4}$
Meng Zhai\orcidlink{0000-0003-1207-3787},$^{1}$
Zewen Jiang\orcidlink{0000-0002-0486-5007},$^{4,8}$
Qianyi Zou\orcidlink{0000-0001-5469-6443},$^{4,8}$
Yujie Peng,$^{4,8}$
\newauthor
Xuedong Gu,$^{6,4}$
Qian Chen$^{9,4}$\\
$^{1}$Chinese Academy of Sciences South America Center for Astronomy, National Astronomical Observatories, Chinese Academy of Sciences, Beijing 100101, China\\
$^{2}$School of Physics and Astronomy, Sun Yat-sen University, Zhuhai 519082, China\\
$^{3}$Center of CSST in the great bay area, Sun Yat-sen University, Zhuhai 519082, China\\
$^{4}$CAS Key Laboratory of Optical Astronomy, National Astronomical Observatories, Chinese Academy of Sciences, Beijing 100101, China\\
$^{5}$Instituto de Astrof\'isica de Canarias (IAC), V\'ia L\'actea s/n, 38205 La Laguna, Tenerife, Spain\\
$^{6}$Departamento de Astrof\'isica, Universidad de La Laguna (ULL), C/ Padre Herrera, 38206 La Laguna, Tenerife, Spain\\
$^{7}$School of Physics and Optoelectronic Engineering, Hainan University, Haikou 570228, China\\
$^{8}$University of Chinese Academy of Sciences, Beijing 100049, China\\
$^{9}$School of Physics and Astronomy, China West Normal University, ShiDa Road, Nanchong 637002, China\\}

\date{Accepted XXX. Received YYY; in original form ZZZ}

\pubyear{2025}
\begin{document}
\label{firstpage}
\pagerange{\pageref{firstpage}--\pageref{lastpage}}
\maketitle

\begin{abstract}
Planets residing within the hot-Neptune Desert are rare, and studying their atmospheres can provide valuable insights into their formation and evolutionary processes. 
We present the atmospheric characterization of the first known ultra-hot Neptune, LTT-9779~b, using transmission spectroscopic observations obtained with the HST/WFC3 G141 and G102 grisms. 
Using the \texttt{Iraclis} pipeline and \texttt{TauREx3} retrieval code, 
we find that LTT-9779~b likely possesses a H/He-dominated primary atmosphere with an opaque aerosol layer and  the pure cloudy, flat-line model is rejected with approximately 2.7-$\sigma$ confidence. 
Although we do not find conclusive evidence supporting the presence of any molecular species, we place 95\% confidence level upper limits on the volume mixing ratios (VMRs) of hydroxyl radical (OH) and iron hydride (FeH) at $7.18\times10^{-2}$ and $1.52\times10^{-8}$, respectively.
Notably, the retrieval results are inconsistent with predictions from equilibrium chemistry models, which favor higher $\rm H_2O$ abundances over $\rm OH$. This discrepancy suggests that disequilibrium processes, such as photochemistry or vertical mixing, may have altered the atmospheric composition.
Comparisons between HST, Spitzer and JWST data reveal no evidence of temporal variations in the atmospheric composition of the terminator region.
Our results highlight the need for higher-resolution spectroscopy and secondary eclipse observations to resolve LTT-9779 b’s temperature-pressure (T-P) profile and chemical inventory definitively.

\end{abstract}
%%%%%%%%%%%%%%%%%%%%%%%%%%%%%%%%%%%%%%%%%%%%%%%%%

%%%%%%%%%%%%%%%%% BODY OF PAPER %%%%%%%%%%%%%%%%%%

%%%%%%%%%%%%%%%%%%%%%%%%%%%%%%%%%%%%%%%%%%%%%%%%%%%%%%%%%%%%%%%%%%%%%%%%%%%%%%%%%%%%%
\section{Introduction} 
\label{sec:intro}
The discovery of planets with extremely short orbital periods (P~$\textless$~1~day), known as ultra-short period (USP) planets, was among the unexpected breakthroughs in exoplanetary science \citep{2014ApJ...787...47S,2018NewAR..83...37W}.
Previously identified USP planets are typically hot Jupiters, with radii exceeding 10~$\rm R_{\oplus}$, or rocky planets, with radii smaller than 2~$\rm R_{\oplus}$.
There is a notable scarcity of short-period planets with radii between 2 $\rm R_{\oplus}$ and 10 $\rm R_{\oplus}$, a phenomenon known as the `Hot-Neptune Desert' \citep{2016A&A...589A..75M}.
Photoevaporation models suggest that Neptune-sized planets within this desert struggle to retain substantial H/He-dominated envelopes due to intense stellar irradiation \citep{Owen12, Lopez17}.
This theory is further supported by comparative studies of young planetary systems and their more mature counterparts \citep{Burt20, Dai23, Fang23}.
Investigating the atmospheres of planets that reside in the hot-Neptune Desert provides crucial information on the mechanisms that allow these planets to endure such extreme environments \citep{2024ApJ...962L..20R}.

In many aspects, ultra-hot Neptunes are similar to hot Jupiters \citep{2021AJ....162...62D}.
Both types of planets have masses above 10~$M_\oplus$, the threshold for runaway gas accretion, and possess H/He or other volatile envelopes.
They also tend to have no neighboring planets in their systems, and they orbit stars with higher metallicities.
Extensive studies have been conducted on the atmospheres of hot Jupiters, both individually and as a population \citep[e.g.,][]{2013Sci...339.1398K, 2014Sci...346..838S, 2023A&A...675A..62J, 2024A&A...682A..73J}.
\citet{2018AJ....155..156T} analyzed the transmission spectra of 30 hot Jupiters obtained with the Hubble Space Telescope's Wide Field Camera 3 (WFC3) and found that the detectability of atmospheres depends on the planetary radius, suggesting the importance of planetary gravity on the atmospheres of hot Jupiters. 
They also detected TiO and VO absorption features in WASP-76~b and WASP-121~b. 
By examining the emission spectra of 25 hot and ultra-hot Jupiters obtained by HST and Spitzer, \citet{2022ApJS..260....3C} revealed a link between the temperature structure of the planets and the abundance of optical absorbers in the atmosphere.
\citet{Edwards23} studied a population of 70 gaseous planets and found evidence for the thermal dissociation of $\rm H_2$ and water via H$^{-}$ opacity in the hotter planets of the sample.

Discovered by the Transiting Exoplanet Survey Satellite \citep[TESS;][]{2015JATIS...1a4003R}, LTT-9779~b is the first ultra-hot Neptune ever found and one of the few planets residing in the hot-Neptune Desert \citep{2020NatAs...4.1148J,2020ApJ...903L...6D}.
It has a radius of 4.6~$\rm R_\oplus$, a mass of 29~$\rm M_\oplus$, and an orbital period of 0.79~days around a bright Sun-like star.
The mean density of LTT-9779~b is similar to that of Neptune.
Due to the intense irradiation from its host star, LTT-9779~b has a high equilibrium temperature.

Using a 1D thermal evolution model to explain the mass-radius relation of LTT-9779~b, \citet{2020NatAs...4.1148J} found that LTT-9779~b needs a H/He envelope that constitutes about 9$\%$ of its total mass.
The non-detection of escaping helium from the atmosphere indicates that it may have weakened evaporation \citep{2024ApJ...962L..19V}. 

\citet{2020ApJ...903L...6D} analyzed the combined observations of the secondary eclipses of LTT-9779~b from Spitzer and TESS. 
They found that the data can be explained by an atmospheric model with CO and CO$_2$ in its atmosphere. 
CHEOPS observations presented by \citet{2023A&A...675A..81H} indicated that LTT-9779~b has a high albedo and a high metallicity atmosphere,  which is consistent with recent studies \citep{2025MNRAS.tmp..375R, 2025A&A...695A..26R, 2025NatAs.tmp...56C}. 
\citet{2023AJ....166..158E} analyzed transmission spectra from HST/WFC3 and found evidence for the presence of $\rm H_2O$, $\rm CO_2$ and FeH.
Using data from JWST/NIRISS, \citet{2024ApJ...962L..20R} found a muted spectrum in LTT-9779~b. 
\citet{Fernandez24} used observations from XMM-Newton to measure an X-ray luminosity that is a factor of 15 lower than expected, supporting the idea that LTT-9779~b could have survived in the `hot-Neptune Desert' due to an unusually weak XUV irradiation history. 

Following our previous atmospheric studies of exoplanets using HST/WFC3 \citep{Zhou20, Zhou23}, we re-analyze the earlier dataset from \citet{2023AJ....166..158E} and investigate the atmosphere of LTT-9779~b in this study by combining the transmission spectral data obtained with the G141 and G102 grisms of HST/WFC3. 
The rest of this paper is outlined below.
Firstly, we describe our data reduction process using \texttt{Iraclis} in Section~\ref{sec:analysis}. 
Then we retrieve the atmospheric properties of LTT-9779~b using \texttt{TauREx3} in Section~\ref{sec:retrieve_method} and present our atmospheric retrieval results in Section~\ref{sec:results}.
In Section~\ref{sec:discussion}, we discuss several aspects of the retrieval results for this planet. 
Finally, we give our conclusions in Section~\ref{sec:conclusions}.

\section{Observation and Data Reduction} \label{sec:analysis}

We downloaded the raw spatially scanned spectroscopic observation data of LTT-9779~b from the public Mikulski Archive for Space Telescopes (\citealt{2020hst..prop16457E}; HST proposal GO-16457, PI: Billy Edwards). 
There is a single transit observation using the G102 and G141 grisms. 
Data were obtained on June 2, 2022, with HST/WFC3 G102 grism (0.8-1.1~$\upmu$m) and on June 11, 2021, with HST/WFC3 G141 grism (1.1-1.7~$\upmu$m). 
The observations were taken using the SPARS10 sequence and the GRISM256 aperture, with an exposure time of 103.129~s per exposure. 
The scan rate was set to 0.18~$\rm ''/s$ with a total scan length of 19.7~$''$.

We applied the open-source pipeline \texttt{Iraclis} \footnote{\url{https://github.com/ucl-exoplanets/Iraclis}} to reduce transit observations. 
\texttt{Iraclis} is designed to analyze the spatially scanned spectral data of HST/WFC3 (\citealp{2016ApJ...832..202T, 2016ApJ...820...99T}). 

The reduction of raw spectroscopic data includes zero-read subtraction, correction for reference pixel and nonlinearity, dark current subtraction, gain conversion, sky background subtraction, correction for bad pixel and cosmic rays \citep{2016ApJ...832..202T,2016ApJ...820...99T,2018AJ....155..156T}.
From the reduced spectroscopic data, we extracted the white and spectral light curves of the transiting exoplanet LTT-9779~b.
The white light curves are generated by integrating the spectrum throughout the wavelength range, which is 0.8$-$1.1~$\upmu$m for the WFC3/G102 band and 1.1$-$1.7~$\upmu$m for the WFC3/G141 band.

To avoid the stronger wavelength-dependent ramp, we discarded the first orbit of each visit in our analysis.
For observations using the G102 grism, we also discarded the first scanning spectroscopic image in each orbit. 
Position shifts, caused by the instrument motion in scanning mode, are estimated against the first image of each visit and are displayed in Figure~\ref{diagnostics}.
We found that the horizontal and vertical shifts are smaller than 0.1~pixel during each visit, which will not significantly affect the precision of transmission spectral measurement. 
We also display the raw white light curves and sky ratios in Figure~\ref{diagnostics}. 

We calculated the limb-darkening coefficients with the non-linear model \citep{2000A&A...363.1081C} using the Atlas stellar model \citep{1970SAOSR.309.....K,2011MNRAS.413.1515H}.
The stellar, planetary, and transit parameters of LTT-9779~b are taken from \citet{2020NatAs...4.1148J}, and are summarized in Table~\ref{parameters}.
During the white-light curve fitting process, the mid-transit time and the planetary radius to stellar radius ratio ($\rm R_p/R_*$) are set as free parameters. 
Other transit, planetary, and stellar parameters, such as orbital inclination, the semi-major axis to stellar radius ratio ($a/R_*$), etc., are fixed.
The parametric model that we use to describe the systematics of the HST/WFC3 light curve is the same as that used in \citet{Zhou20, Zhou23}.

\begin{figure*}
\includegraphics[width=\textwidth]{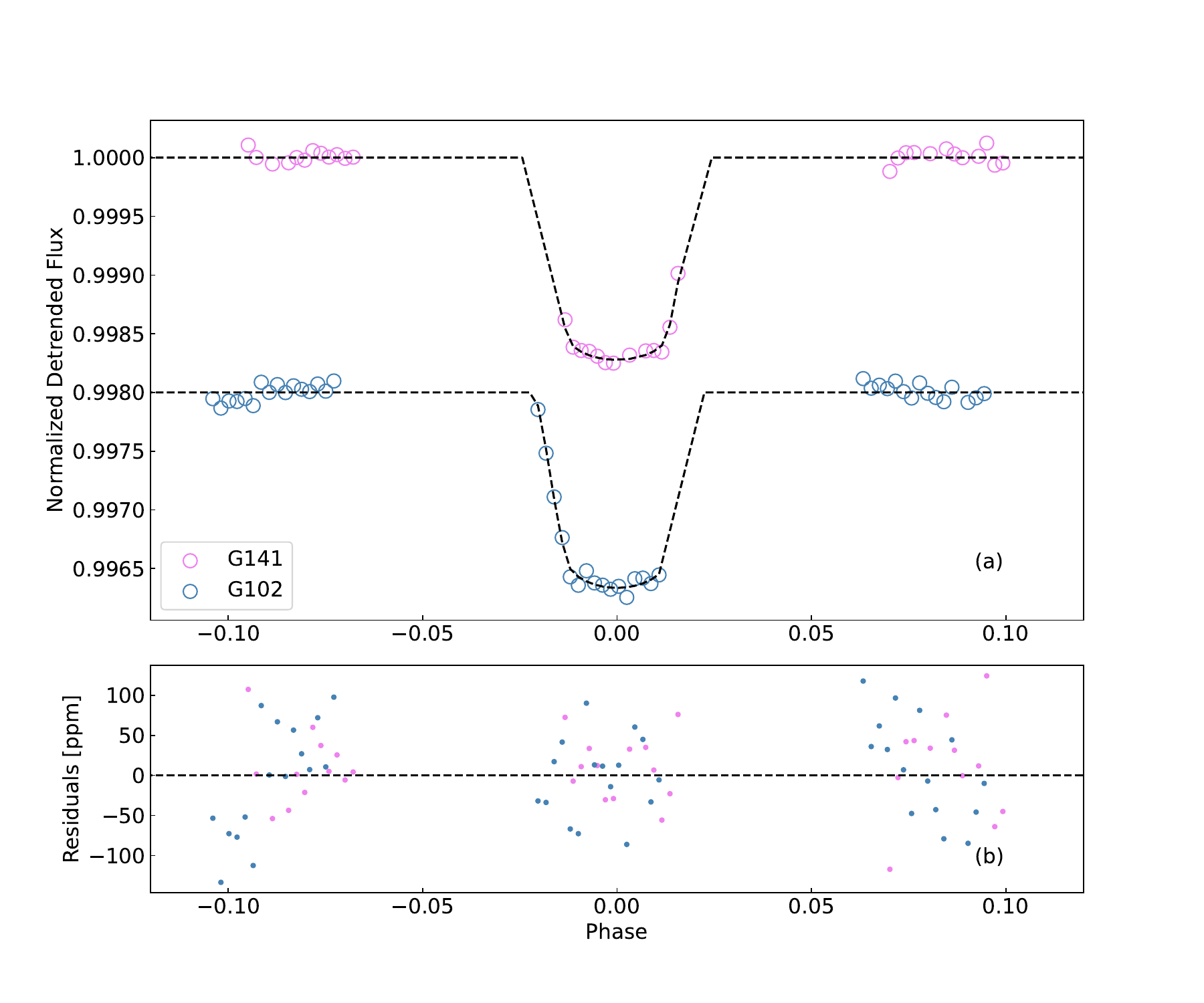}
\caption{Top panel: White light curves and best-fit transit models of LTT-9779~b from WFC3/G141 and G102 grism. Bottom panel: fitting residuals. An offset has been applied in the top panel for display purpose.}
\label{lc}
\end{figure*}

There is a difference in how the short-term ramp is treated between this work and \citet{2023AJ....166..158E}. 
In \citet{2023AJ....166..158E}, separate ramp coefficients ($\rm for_{b1}$, $\rm for_{b2}$) were used for the first orbit, whereas in our work a single set of coefficients is applied. 
In \citet{2023AJ....166..158E}, there is a correlation between the radius ratio of $\rm R_p/R_*$ and the systematic parameters, suggesting challenges in determining absolute transit depths. 
We present the white light curve correlations of different parameters in this work in Figure~\ref{white_correlations}. 
We can see that there is no correlation between them, indicating more precise and reliable absolute transit depths.
We present the white light curves, the best-fit transit models, and the residuals in Figure~\ref{lc}, and the best-fit mid-transit time value presented in Table~\ref{parameters} will be used in our future transit timing variation studies \citep{Wang24}.

\begin{figure*}
\includegraphics[width=\textwidth]{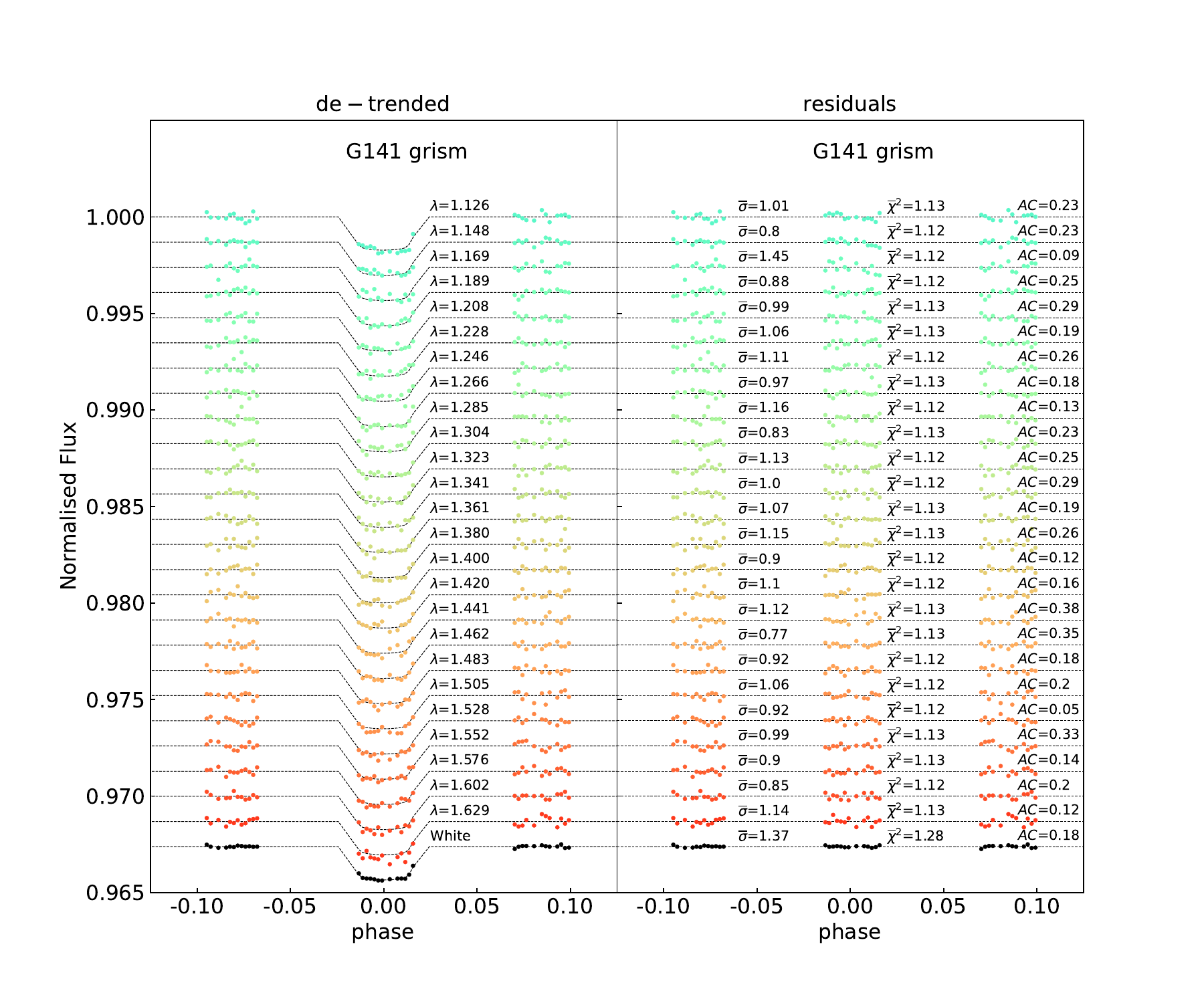}
\caption{Spectral transit light curves of LTT-9779~b from the observations of HST/WFC3 G141 grism. Left panel: the de-trended spectral light curves with the best-fit transit model plotted for each wavelength bin. 
Right panel: the fitting residuals of the spectral light curves. We have also calculated the standard deviation with respect to the photon noise ($\overline\sigma$), the reduced chi-square of residuals from the fitting ($\overline\chi^2$) and the auto-correlation (AC) during the transit. Offsets have been applied to the light curves for display purpose.}
\label{all_lc_g141}
\end{figure*}

\begin{figure*}
\includegraphics[width=\textwidth]{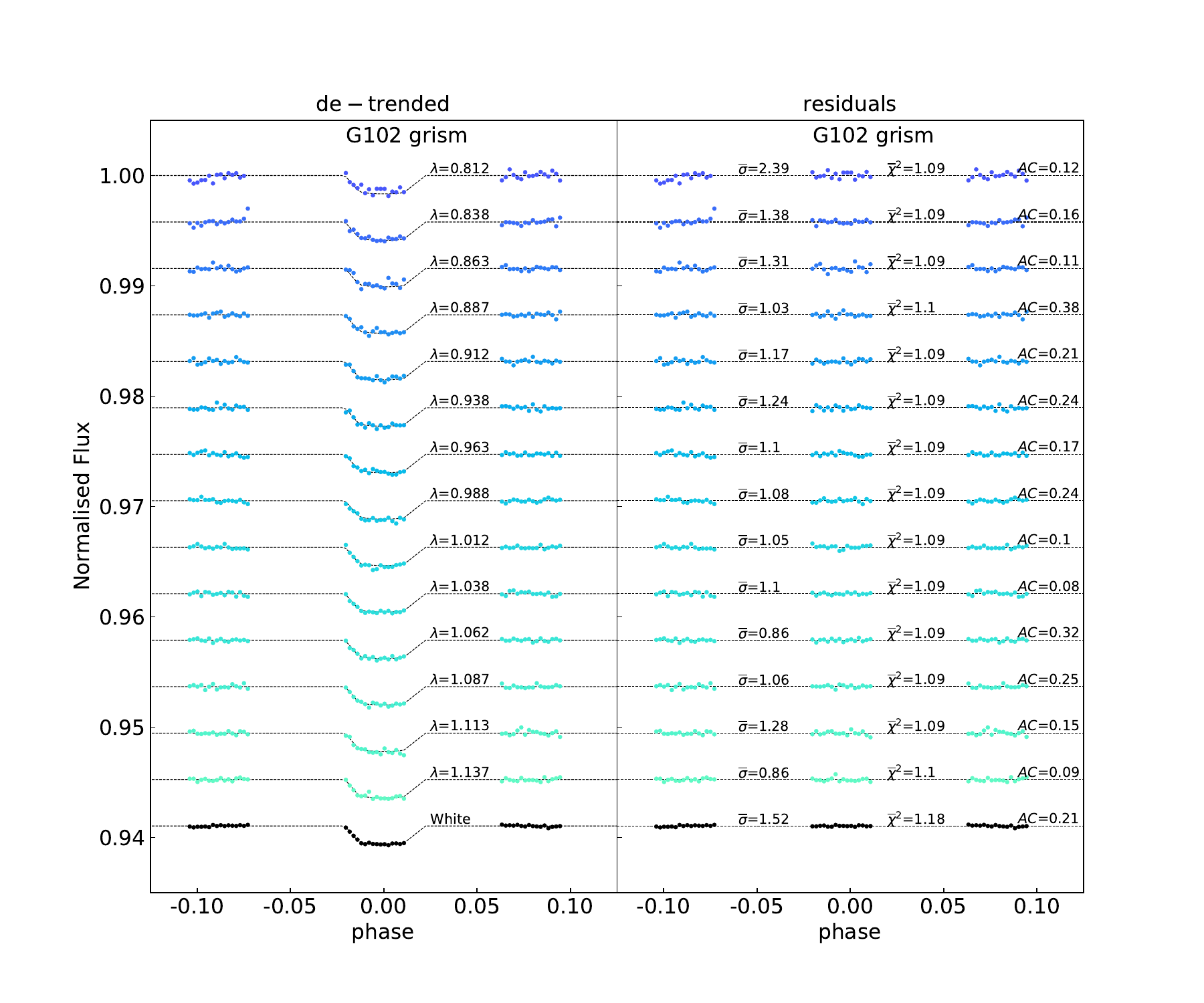}
\caption{Similar with Figure~\ref{all_lc_g141}, but for observations using G102 grism. }
\label{all_lc_g102}
\end{figure*}

Next, we extracted the spectral light curves by adopting the default--high binning setting in \texttt{Iraclis}, which corresponds to a resolving power of 70 at 1.4~$\upmu$m. 
We fit these spectral light curves with a transit model plus a systematic model similar to that used in \citet{2016ApJ...832..202T}.
The mid-transit time derived from the fitting of the white-light curve is fixed in this fitting process, and $\rm R_p/R_*$ is the only free parameter in the transit model.
We present the derived spectral light curves of G141 and G102 grisms in Figure~\ref{all_lc_g141} and Figure~\ref{all_lc_g102}, respectively.

\begin{table}
\centering
\small
\caption{Parameters of LTT-9779~b and its host star, adopted from \citet{2020NatAs...4.1148J}.} \label{parameters}
\resizebox{0.3\textwidth}{!}{
\begin{tabular}[b]{cc}
\hline
\multicolumn{2}{c}{Stellar Parameters}\\
\hline
 $\rm [Fe/H](dex)$  & 0.27 $\pm$ 0.03\\
 $T_{\rm eff}(\rm K)$ & 5443 $^{+14}_{-13}$\\
 $M_{\rm *}(M_{\odot})$ & 0.77 $^{+0.29}_{-0.21}$\\
 $R_{\rm *}(R_\odot)$ & 0.949 $\pm$ 0.006\\
 $\mathrm {log}\,g_*(\rm cgs)$ & 4.35 $^{+0.16}_{-0.12}$\\
\hline
\multicolumn{2}{c}{Planetary  Parameters}\\
\hline
$T_{\rm eq}(\rm K)$ & 1978 $\pm$ 19 \\
$M_{\rm p}(M_\oplus)$ & 29.32 $^{+0.78}_{-0.81}$\\
$R_{\rm p}(R_\oplus)$ & 4.72 $\pm$ 0.23\\
\hline
\multicolumn{2}{c}{Transit  Parameters}\\
\hline
$T_{\rm 0}(\rm BJD)$ & 2458354.21430 $\pm$ 0.00025 \\
$P(\rm days)$ & 0.7920520 $\pm$ 0.0000093\\
$R_{\rm p}/R_{\rm *}$ & 0.0455 $^{+0.0022}_{-0.0017}$ \\
$a(\rm au)$ & 0.01679 $^{+0.00014}_{-0.00012}$ \\
$a/R_*$ & 3.877 $^{+0.090}_{-0.091}$ \\
$i(\rm deg)$ & 76.39 $\pm$ 0.43 \\
\hline
\end{tabular}}
\end{table}

%white light curve figures

%all light curves of spectrum of G141
%G102

\section{Atmospheric Retrieval} \label{sec:retrieve}
In this section, we perform both free chemistry and equilibrium chemistry retrievals using combined observations from the HST/WFC3 G141 and G102 grisms, and present the corresponding results.

\subsection{Retrieval Method} \label{sec:retrieve_method}
We utilized the publicly available code \texttt{TauREx3}\footnote{\url{https://github.com/ucl-exoplanets/TauREx3_public}} \citep{2021ApJ...917...37A, 2015ApJ...813...13W,  2015ApJ...802..107W} to retrieve the atmospheric parameters of LTT-9779~b from its extracted transmission spectrum. 
\texttt{TauREx3} uses a Bayesian atmospheric retrieval framework and the nested sampling algorithm \texttt{Multinest} \citep{2014A&A...564A.125B} to explore the planetary atmospheric parameter space and determine the best-fit atmospheric model for the observed spectrum. 
For our retrievals, we configured \texttt{Multinest} with 750 live points and set the evidence tolerance to 0.5.

In the 1D atmospheric model, the altitude is parameterized using a grid of 100 layers, with pressure sampled uniformly in logarithmic space, ranging from  $\rm 10^{-4}$~Pa at the top of the atmosphere to $\rm 10^6$~Pa at the planetary surface. 
Given the limited wavelength coverage of the HST spectrum, an isothermal T--P profile is adopted.
The prior for the equilibrium temperature is defined within the range of 0.6 $T_{\rm eq}$ to 1.6 $T_{\rm eq}$, where $T_{\rm eq} = 1978 K$ is taken from Table~\ref{parameters}. 
Similarly, the prior for the planetary radius is set to a uniform range from 0.6 $R_{\rm p}$ to 1.6 $R_{\rm p}$.

In our analysis, we included $\rm H_2$ and $\rm He$ as filling gases, fixing the $\rm He/H_2$ ratio to the solar value of 0.17. 
We adopted line lists for active chemical gases from ExoMol \citep{2016JMoSp.327...73T,2021A&A...646A..21C}, HITEMP \citep{2010JQSRT.111.2139R} and HITRAN \citep{1987ApOpt..26.4058R},
including $\rm HCN$ \citep{2014MNRAS.437.1828B}, $\rm CH_4$ \citep{2014MNRAS.440.1649Y}, $\rm NH_3$ \citep{2019MNRAS.490.4638C}, $\rm CO$ \citep{2015ApJS..216...15L}, $\rm CO_2$ \citep{2010JQSRT.111.2139R}, $\rm H_2O$ \citep{2018MNRAS.480.2597P}, $\rm TiO$ \citep{2019MNRAS.488.2836M}, $\rm MgH$ \citep{2013MNRAS.432.2043G}, $\rm FeH$ \citep{2020JQSRT.24006687B}, $\rm AlH$ \citep{2018MNRAS.479.1401Y}, $\rm OH$ \citep{2018JQSRT.217..416Y}, $\rm C_2H_4$ \citep{2018MNRAS.478.3220M}, $\rm CP$ \citep{2014JQSRT.138..107R}, and $\rm VO$ \citep{2016MNRAS.463..771M} etc.
The inclusion of these molecules in the retrieval process is motivated by several considerations \citep{2021A&A...646A..17B}: predictions from thermal equilibrium models regarding the presence and abundances of various species \citep{2018A&A...614A...1W}, distinct absorption features associated with nitrogen- and carbon-bearing molecules, metal hydrides, and oxides \citep{2007ApJS..168..140S}, and indicators of photochemical processes inferred from species such as OH and $\rm C_2H_2$ \citep{2010ApJ...717..496L, 2019ApJ...877..109K}. 
The complete list of line lists used in our analysis is presented in Table~\ref{lines_taurex}. 
Additionally, the planetary atmospheric models include Rayleigh scattering and collision-induced absorption (CIA) of $\rm H_2$--$\rm H_2$ \citep{doi:10.1021/jp109441f,2018ApJS..235...24F} and $\rm H_2$--$\rm He$ \citep{doi:10.1063/1.3676405}.
Gray clouds are included in the models with a top--pressure ranging from $\rm 10^{-4}$ to $\rm 10^{6}$~Pa, following a log-uniform prior\citep{2013ApJ...778...97L} .

\begin{table}
\centering
\caption{Molecular line lists used in this analysis.}
\label{lines_taurex}
\resizebox{0.5\textwidth}{!}{
\begin{tabular}[b]{l|l|l}
\hline
Molecule & Database & Reference \\
\hline
OH & MoLLIST & \citet{2018JQSRT.217..416Y} \\
$\rm H_2O$ & ExoMol & \citet{2018MNRAS.480.2597P}\\
AlH & ExoMol & \citet{2018MNRAS.479.1401Y} \\
AlO & ExoMol & \citet{2015MNRAS.449.3613P}\\
$\rm C_2H_2$ & ExoMol & \citet{2020MNRAS.493.1531C}\\
$\rm C_2H_4$ & ExoMol & \citet{2018MNRAS.478.3220M}\\
CaH & MOLLIST & \citet{2012JQSRT.113...67L, 2020JQSRT.24006687B}\\
$\rm CH_4$ & ExoMol & \citet{2014MNRAS.440.1649Y}\\
CN & MOLLIST & \citet{2014ApJS..210...23B} \\
$\rm CO_2$ & HITEMP & \citet{2010JQSRT.111.2139R}\\
CO & ExoMol & \citet{2015ApJS..216...15L}\\
CP & MOLLIST & \citet{2014JQSRT.138..107R}\\
CrH & MOLLIST & \citet{2002ApJ...577..986B, 2020JQSRT.24006687B}\\
FeH & MOLLIST & \citet{2020JQSRT.24006687B}\\
$\rm H_2CO$ & ExoMol & \citet{2015MNRAS.448.1704A}\\
HCN & ExoMol & \citet{2014MNRAS.437.1828B}\\
MgH & MOLLIST & \citet{2013MNRAS.432.2043G, 2020JQSRT.24006687B}\\
MgO & ExoMol & \citet{2019MNRAS.486.2351L}\\
$\rm NH_3$ & ExoMol & \citet{2019MNRAS.490.4638C}\\
ScH & ExoMol & \citet{2015MolPh.113.1998L}\\
% ScH & ExoMol & \citet{2015MolPh.113.1998L, 2021A&A...646A..21C}\\
TiH & MOLLIST & \citet{2005ApJ...624..988B}\\
TiO & ExoMol & \citet{2019MNRAS.488.2836M}\\
VO & ExoMol & \citet{2016MNRAS.463..771M}\\
\hline
\end{tabular}}
\end{table}

Given the limited number of data points and the narrow spectral coverage, we cannot include too many chemical components simultaneously. Therefore, we adopt a bottom-up approach, which is also effective in quantifying the significance of individual molecules \citep{2021A&A...646A..17B}.

To search for the presence of chemical gases in the transmission spectrum, we initially performed 25 retrievals, each assuming a different atmospheric model.
These include 23 models in which a single molecule from Table~\ref{lines_taurex} is considered as a chemical gas, one pure cloudy flat-line model that includes only gray clouds, and one model incorporating gray clouds, Rayleigh scattering, and collision-induced absorption (CIA).

Next, we performed retrievals assuming atmospheric models with two different chemical gases selected from the 23 chemical gases in Table~\ref{lines_taurex}. 
Bayesian evidence indicates that OH is the most favorable single chemical gas, and the combination of OH and FeH is the most favorable pair among those tested.
Subsequently, we performed retrievals with three molecules, including OH, FeH, and one other species, to assess the presence of additional potential atmospheric constituents. 
We find that none of the other species are favored at a statistically significant level.
Thus, we obtain the final best-fit atmospheric model by incorporating only OH and FeH as chemical gases. This model serves as the baseline for evaluating the detection significance of each molecule.
To evaluate the detection significance of OH and FeH, we systematically excluded one molecule at a time from this final model.

In all retrievals, the volume mixing ratios (VMRs) of the chemical gases are allowed to vary between $\rm 10^{-12}$ and $\rm 10^{-1}$. 
After completing these retrievals, we used the equilibrium chemistry code \texttt{FastChem} \citep{2018MNRAS.479..865S,2022MNRAS.517.4070S}, integrated into \texttt{TauREx3} as a plugin \citep{2022ApJ...932..123A}, to investigate the atmospheric metallicity and C/O ratio of LTT-9779~b. 
The elements included in the equilibrium chemistry model are H, He, C, N, O, Fe, Mg, Ti, Ca, Al, Sc, Cr and V.
The priors for the C/O ratio and metallicity are consistent with those used by \citet{2023AJ....166..158E}, with the C/O ratio following a linear-uniform distribution from 0.001 to 2, and the metallicity following a log-uniform distribution from -4 to +4. 
\subsection{Retrieval Results} \label{sec:results}
In this section, we present the atmospheric retrieval results based on the models outlined in Section~\ref{sec:retrieve_method}.
The statistical strength of atmospheric detections is assessed using the logarithmic Bayes factor ($\rm \Delta ln(E)$).

We evaluated the models against a baseline pure cloudy flat-line model, defining the Bayes factor as the difference in Bayesian log evidence ($\rm ln(E)$) between the model under evaluation and the baseline model. 
The statistical strength of a detection is classified as strong when the Bayes factor exceeds 5.0 (approximately 3.6~$\sigma$ according to the conversion of \citealt{2013ApJ...778..153B}), moderate when it is between 2.5 and 5.0 (or between 2.7 and 3.6~$\sigma$), and weak when it falls below 2.5 (or below 2.7~$\sigma$).
A summary of the statistical results, including the $\rm \Delta ln(E)$ and the corresponding n-$\sigma$ values, for all retrievals is provided in Table~\ref{retrieval_results_constant}.

\begin{figure}
\includegraphics[width=0.5\textwidth]{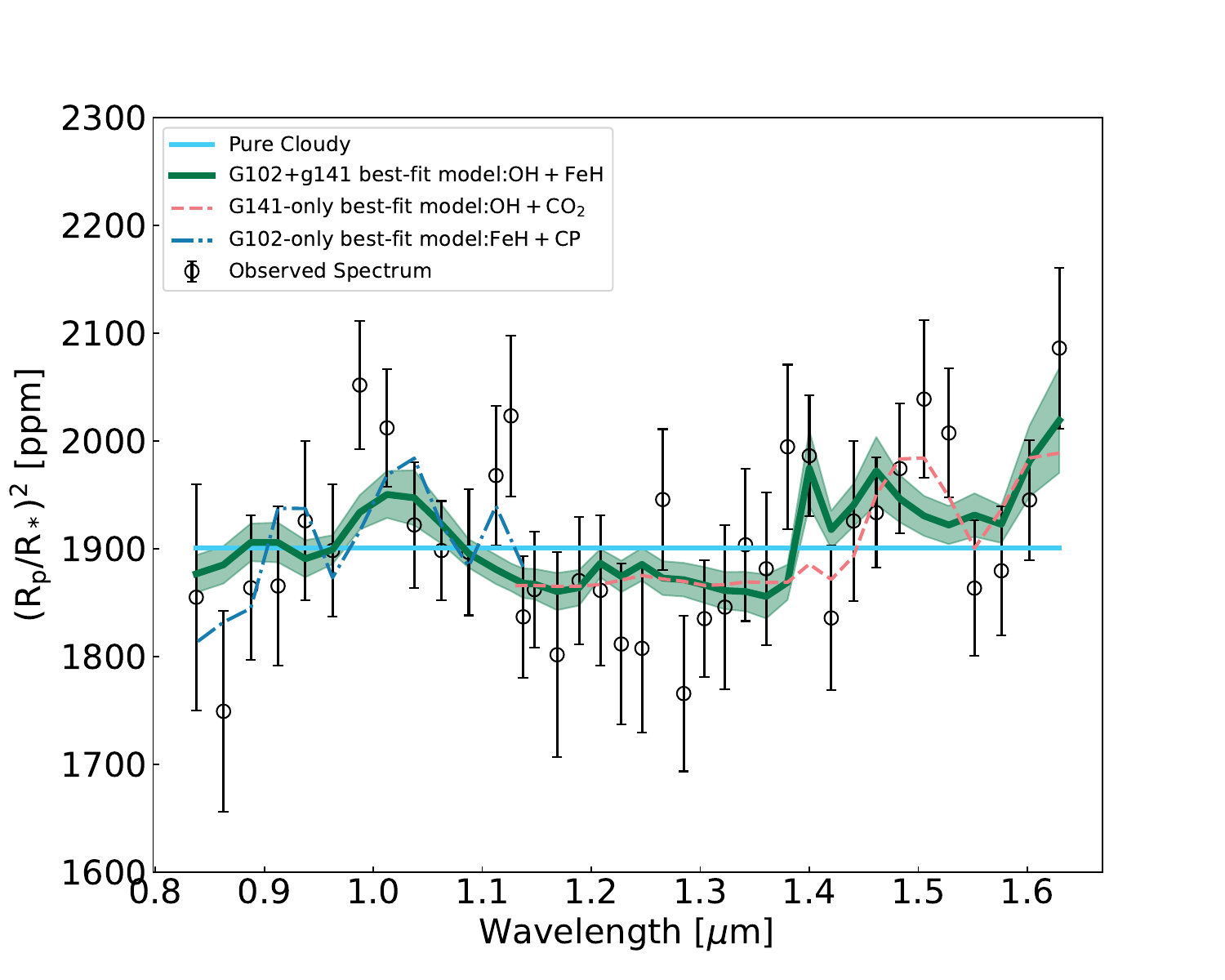}
\caption{Transmission spectra of different atmospheric models for LTT-9779~b using observations from G102 and G141 grisms.}
\label{com_models_g141_g102}
\end{figure}

\begin{figure*}
\includegraphics[width=1.0\textwidth]{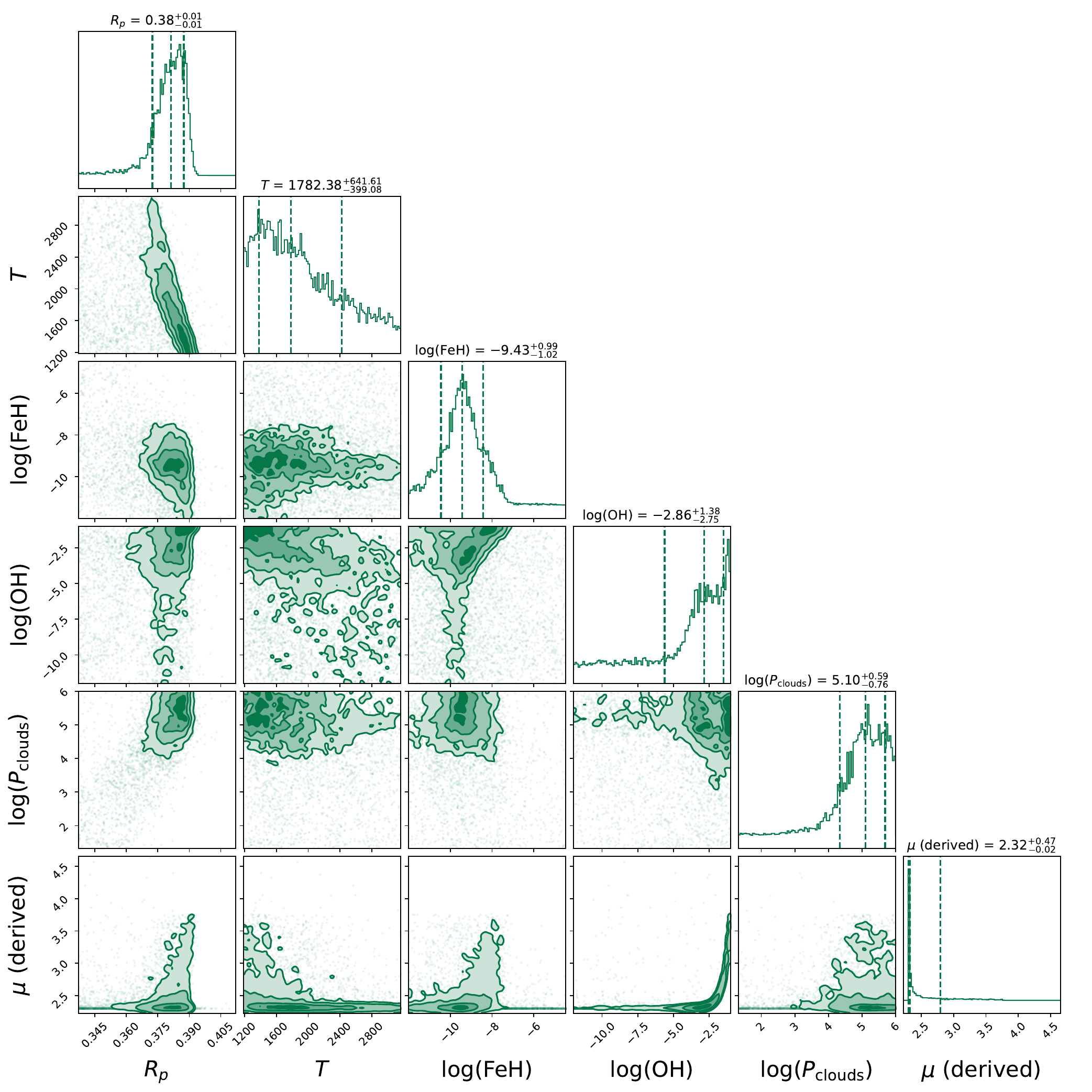}
\caption{Posterior distributions of the atmospheric retrieval parameters for the final model (primary cloud model with OH and FeH as chemical gas) for LTT-9779~b, based on observations from the HST/WFC3 G102 and G141 prisms.}
\label{posterior_g102_g141}
\end{figure*}

\begin{table}
\centering
\caption{Retrieval parameters for the best-fit free chemistry and equilibrium chemistry models. 
The third column shows the 68\% confidence intervals (lower and upper limits) derived from the HST/WFC3 G102 and G141 data.
The final column provides the retrieval results from the combined analysis of the HST/WFC3 data from this work, including additional observations from JWST \citep{2024ApJ...962L..20R}.}
\label{retrieval_pars}
\begin{threeparttable}
\resizebox{0.5\textwidth}{!}{
\begin{tabular}[b]{llll}
\hline
\multicolumn{4}{c}{Free Chemistry}\\
 &Prior& Posterior (HST/WFC3) & Posterior (+JWST)\\
 $\rm R_p~(R_{Jup})$ & $\rm 0.2565\sim0.6736$ & [0.37, 0.39] & [0.37, 0.38]\\
 T~(K) & $\rm 1186\sim3165$ & [1383.30, 2423.99] & [1258.85, 2095.54]
\\
 $\rm log_{10}(P)~(Pa)$ & $-4\sim6$ & [4.34, 5.69] & [4.21, 4.89]\\
 $\rm log_{10}(OH)$ & $-12\sim-1$ & [-5.61, -1.48] & [-4.23, -2.92]\\
 $\rm log_{10}(FeH)$ & $-12\sim-1$ & [-10.45, -8.44] & [-11.48, -9.89]\\
\hline
\multicolumn{4}{c}{Equilibrium Chemistry}\\
 &Prior& Posterior(HST/WFC3) & Posterior(+JWST)\\
 $\rm R_p~(R_{Jup})$ & $\rm 0.2565\sim0.6736$ & [0.34, 0.40] & [0.38, 0.40]\\
 T~(K) & $\rm 1186\sim3165$ & [1444.17, 2555.93]& [1514.33, 1956.30]\\
 $\rm log_{10}(P)~(Pa)$ & $-4\sim6$ & [-0.57, 4.91] & [2.14, 5.23]\\
 $\rm C/O $ & $0.001\sim2$ & [0.34, 1.62] & [0.08, 0.52]\\
 $\rm log_{10}(Z)~(Z_\odot)$ & $-4\sim4$ & [-3.11, 3.06] & [2.21, 2.78]\\
\hline
\end{tabular}}
\end{threeparttable}
\end{table}
From Table~\ref{retrieval_results_constant}, we conclude that the pure cloudy model can be rejected at a confidence level of $\sim$ 2.7~$\sigma$. 
The final best-fit model incorporates OH and FeH as chemical gases within a primary atmosphere, along with a low-altitude cloud deck, as illustrated in Figure~\ref{com_models_g141_g102}. 
The posterior distributions and specific values of the retrieval parameters for the best-fit model are presented in Figure~\ref{posterior_g102_g141} and Table~\ref{retrieval_pars}. 
The retrieved best-fit temperature is $\sim$1782~K, similar with the equilibrium temperature derived using a high albedo value of 0.7 in the study by \citet{2024ApJ...962L..20R}.
The gray cloud's top layer is located at a pressure of roughly 1 bar, suggesting that the clouds are situated deep within the atmosphere.

\begin{table}
\centering
\caption{Bayesian log evidence and reduced chi-square ($\overline{\chi}^2$) values for the final best-fit retrieval and for retrievals excluding one molecule at a time from the final best-fit model.}
\label{retrieval_lge}
\begin{threeparttable}
\resizebox{0.5\textwidth}{!}{
\begin{tabular}[b]{llll}
\hline
\multicolumn{4}{c}{Free Chemistry}\\
 Model&ln(E)& Sigma & $\overline \chi^2$\\
 &&(compared with final model)&\\
 final model & 302.883 & - & 1.07\\
 without OH & 301.426 & -2.3 & 1.16 \\
 without FeH & 302.301 & -1.7 & 1.07\\
\hline
\multicolumn{4}{c}{Chemical Equilibrium}\\
 FastChem & 299.86 & -3.0 & 1.37 \\
\hline
\end{tabular}}
\end{threeparttable}
\end{table}
%boma
We illustrate the contributions of individual species to the final model in Figure~\ref{contributions_hst_full_model}.
From this figure, it is clear that the absorption features in the transmission spectrum between 1.4~$\upmu$m and 1.5~$\upmu$m are primarily due to $\rm OH$.
The features near the red end of the G102 spectrum likely arise from FeH, which shows a significant contribution. 

Table~\ref{retrieval_lge} summarizes the retrieval results, including ln(E) and $\overline{\chi}^2$, for the best-fit final model as well as for models in which either OH or FeH is excluded. 
These results indicate that the free-chemistry final model is the most plausible, and the equilibrium chemistry model can be rejected at a confidence level of 3.0~$\sigma$ when compared with the final model. 
However, the detection of either OH or FeH is statistically insignificant, and the constraints on their abundances are weak, as shown by the corner plots in Figure~\ref{posterior_g102_g141}.

Therefore, atmospheric retrievals of the combined HST/WFC3 G141 and G102 grism observations reveal that LTT-9779 b's atmosphere is best explained by a H/He-dominated primordial envelope with a low-altitude opaque cloud deck. 
Although we cannot confirm the presence of OH and FeH in LTT-9779 b's atmosphere, we obtain upper limits on their VMRs of $7.18\times10^{-2}$ and $1.52\times10^{-8}$, respectively, at the 95\% confidence level.

\section{Discussion}
\label{sec:discussion}
\subsection{Comparison with Previous Works}
\label{sec:compare}

\begin{figure*}
\includegraphics[width=\textwidth]{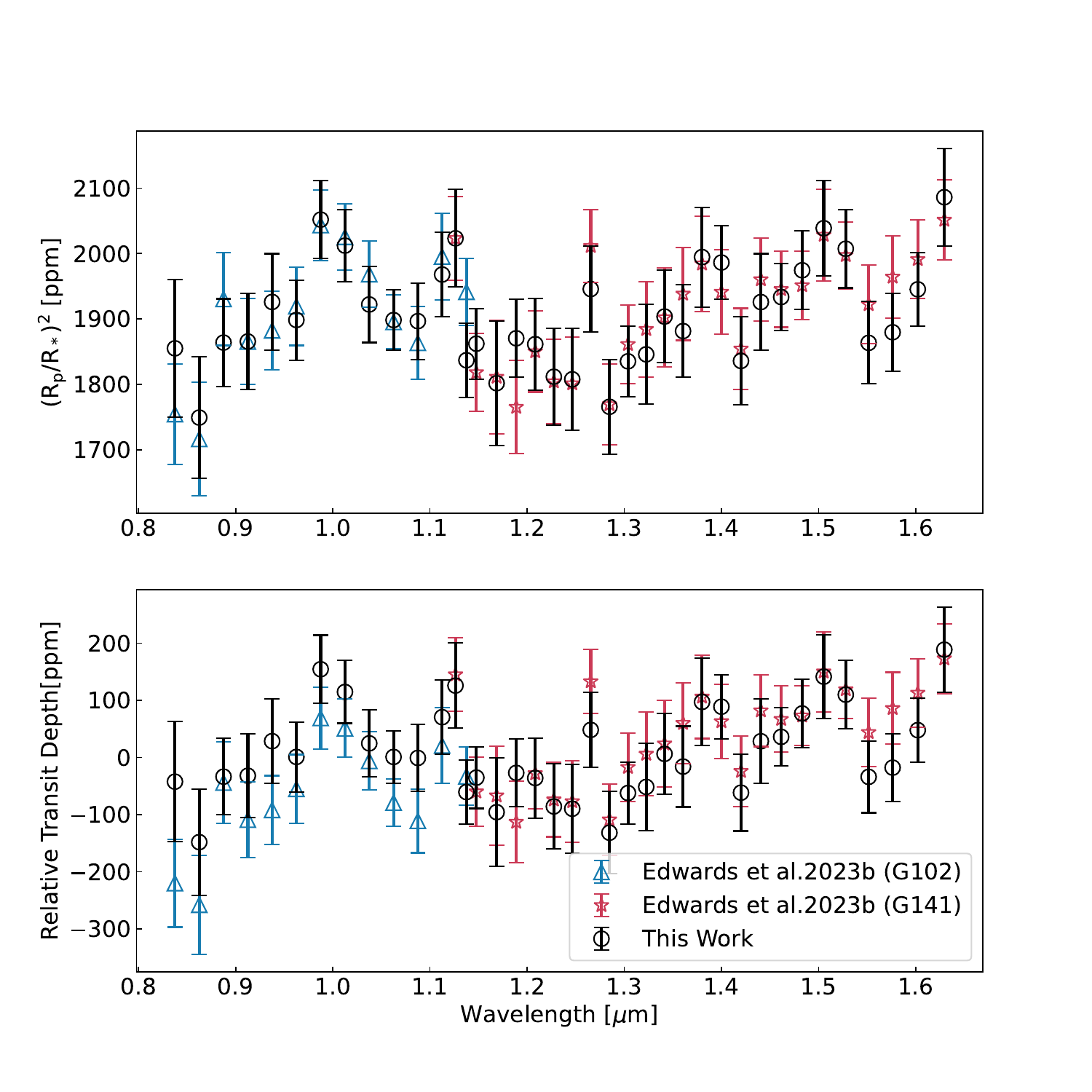}
\caption{Comparison between our transmission spectrum and that from \citet{2023AJ....166..158E}. In the upper panel, the G102 and G141 data from \citet{2023AJ....166..158E} are reduced by offsets of 55.5 ppm and 151.7 ppm, respectively. In the bottom panel, the data show the relative transit depths, which have been normalized to their respective medians. The  median values are 2030.0 ppm for the observations from \citet{2023AJ....166..158E} and 1897.5 ppm for the data presented in this work.}
\label{compare_ours_edwards}
\end{figure*}

LTT-9779~b has enriched a valuable data on the small sample size of sub-Jovian, for which the roles of evolutionary history and birth environment in determining the planets' varied atmospheric compositions have not been well distinguished \citep{2020ApJ...903L...6D}. 
LTT-9779~b is assumed to form from an initially more massive and larger planet.
\citet{2023AJ....166..158E} analyzed the same HST/WFC3 data for LTT-9779~b. 
The trend of the transmission spectral data obtained in this work is mostly consistent with those from \citet{2023AJ....166..158E}, but with an offset of $\sim$ 151.7 ppm for the G141 data, and $\sim$ 55.5 ppm for the G102 data, which are shown in Figure~\ref{compare_ours_edwards}.
These offsets are likely due to differences in the coefficients used to correct the short ramp effect in the first orbit during white light curve fitting (as discussed in Sect.~\ref{sec:analysis}), or variations in the sky background subtraction area during data reduction. 
This constitutes the first key difference between the two studies. 
The Kolmogorov-Smirnov (KS) test that compares the relative transit depths of this work and \citet{2023AJ....166..158E} yields a D-value of 0.16 and a p-value of 0.74, indicating that there are no significant differences between the relative depth results of the two reductions.
Furthermore, we compared our spectrum with that obtained from JWST \citep{2024ApJ...962L..20R}, as shown in Figure~\ref{wfc3-spitzer-jwst}, which supports the reliability of the data reduction presented in this study.

The second major difference between the two studies lies in the retrieval results. 
While \citet{2023AJ....166..158E} identifies evidence for the presence of H$_2$O, CO$_2$, and FeH in their retrieval, 
our retrieval finds only weak evidence for the presence of OH and FeH. 
Furthermore, both retrieval results are different from that reported by \citet{2024ApJ...962L..20R} using the JWST data.

The atmospheric C/O ratio provides critical insights into planet formation processes \citep{2012ApJ...758...36M,2017MNRAS.469.4102M,2020AJ....160..150W}, while metallicity offers valuable clues about the extent of mass loss and its role in preserving the metallicity of a planet's primordial atmosphere.
\citet{2020ApJ...903L...6D} reported evidence of CO with a VMR of $\rm log_{10}$(CO) =$-3^{+1.3}_{-1.7}$~dex by analyzing the secondary eclipse data of LTT-9779~b obtained from Spitzer and TESS. 
However, consistent with other studies that rely on broadband emission spectra of hot-Jupiters, their results highlight a degeneracy between the abundances of CO and CO$_2$ when using only Spitzer observations at 4.5~$\upmu$m. 
They derived a C/O ratio lower than 1, assuming the high temperature of the planet. 
Additionally, \citet{2020ApJ...903L...7C} analyzed the Spitzer phase curve of LTT-9779~b and inferred a supersolar atmospheric metallicity for LTT-9779~b. 
\citet{2023A&A...675A..81H} found that a supersolar metallicity, above 400 times solar, is needed to explain the high albedo derived from the secondary eclipse observation of LTT-9779~b. 
Using a spectrum from JWST/NIRISS GR700XD , \citet{2024ApJ...962L..20R} further constrained the metallicity of LTT-9779~b within a range of 20 to 850 times solar. 
\cite{2025A&A...695A..26R} placed a lower bound of Z $\textgreater$ 180~Z$_\odot$ for the atmospheric metallicity from ESPRESSO observations.

In this work, using spectroscopic observations from the WFC3/G102 and G141 grisms, we are unable to resolve the degeneracy between $\rm CO_2$ and CO, nor can we confirm the detection of either of them.
Additionally, we conducted a free chemistry retrieval combining HST/WFC3 data from this work with the JWST data from \citet{2024ApJ...962L..20R}, yielding similar results. 
The best-fit retrieval parameters and posterior distributions are presented in Table~\ref{retrieval_pars} , Figure~\ref{posterior_g102_g141} and Figure~\ref{posterior_hst_jwst}.

Using HST/WFC3 data, we derive C/O ratio of $\rm C/O=0.91^{+0.71}_{-0.57}$ and a very low metallicity of $\rm log_{10}(Z)=-1.09^{+4.15}_{-2.02}$ from equilibrium chemistry retrievals performed with the \texttt{FastChem} plugin in \texttt{TauREx3}. 
However, as shown in Table~\ref{retrieval_lge}, these equilibrium chemistry retrieval results are not reliable when based solely on HST/WFC3 data.
When combining the HST/WFC3 data with JWST data from \citet{2024ApJ...962L..20R}, the retrieved $\rm C/O$ ratio is $\rm C/O=0.25^{+0.27}_{-0.17}$, while the metallicity increases to $\rm log_{10}(Z)=2.44^{+0.34}_{-0.23}$. 
These values are consistent with the findings of previous studies (\citealp{2020ApJ...903L...7C,2023A&A...675A..81H,2024ApJ...962L..20R, 2025A&A...695A..26R}).

\subsection{Interpretation using Equilibrium Chemistry}

\label{sec:equilibrium chemistry}
Previously, the chemical equilibrium retrievals conducted by \citet{2023AJ....166..158E} did not find a preferable fit to the data. 
Similarly, we find that the atmosphere of LTT-9779~b cannot be explained by equilibrium chemistry. 
Furthermore, the HST/WFC3 wavelength coverage does not provide a strong sensitivity to carbon-bearing species, and as a result, the C/O ratio is poorly constrained \citep{2016ApJ...833..120R,2020AJ....160..260C}.
We apply the open-source chemical equilibrium code \texttt{FastChem} to calculate the chemical composition of the gas phase for a given temperature and pressure under the assumption of chemical equilibrium.

\begin{figure*}
\includegraphics[width=\textwidth]{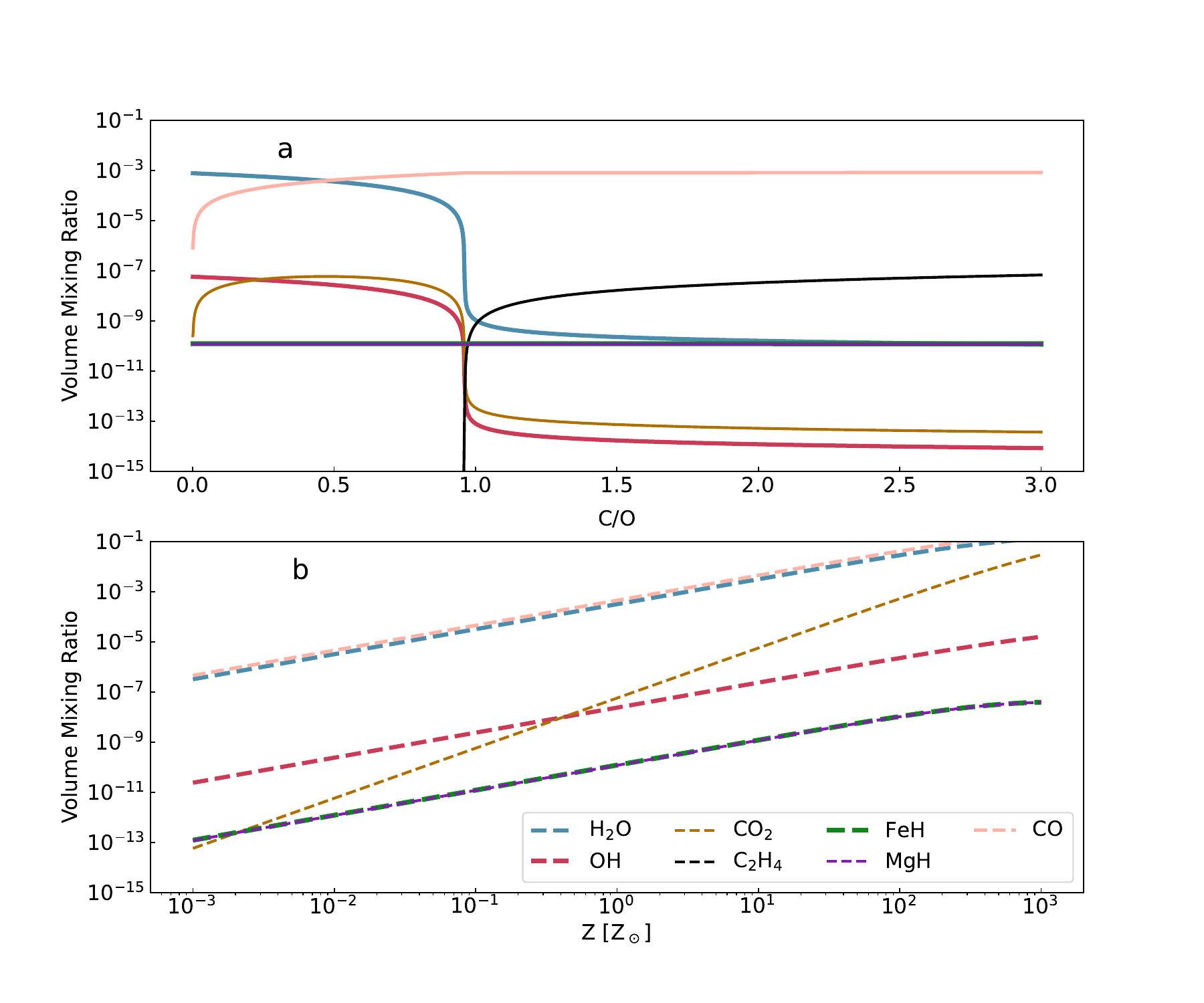}
\caption{Panel (a): The abundance of molecules varies with different C/O ratios, calculated using the equilibrium chemistry model assuming solar metallicity. We assume solar photospheric abundances for all elements except for C and O.
Panel (b): The abundance of molecules varies with different atmospheric metallicities (from 0.001 to 1000 times solar metallicity), calculated using equilibrium chemistry model.}
\label{fastchem_c_o_z}
\end{figure*}

The temperature and pressure are set to 1782~K and 1~mbar, respectively.
We assumed solar photospheric abundances for all elements except C and O \citep{2009ARA&A..47..481A} and explored the variation in the abundances of molecular species with different C/O ratio. 
The results are shown in Figure~\ref{fastchem_c_o_z}.
There is a sudden change in the abundance of most molecules when the C/O ratio increases from $\leq$ 1 to $\geq$ 1.
Hydrocarbon species such as $\rm C_2H_2$, $\rm CH_4$, and HCN become more prominent when C/O $\geq$ 1, while oxide species like $\rm CO_2$, $\rm H_2O$, TiO and VO become more prominent when C/O $\leq$ 1 \citep{2012ApJ...758...36M}. 
In the final model, the upper limits of OH VMR is high, which corresponds to a C/O ratio lower than 1. 
This is consistent with the C/O ratio retrieval results when combining HST and JWST data using the equilibrium chemistry model in Section~\ref{sec:compare}. 

We have also investigated the impact of atmospheric metallicity on the abundance of molecular species, as shown in Figure~\ref{fastchem_c_o_z}. 
The study by \citet{2020ApJ...903L...7C} suggests that LTT-9779~b has super-solar atmospheric metallicity. 
We can see that the abundances of most molecular species increase with increasing metallicity, from sub-solar to super-solar metallicity. 
The molecule of $\rm H_2O$ remains prominent under different metallicity conditions.
Additionally, Figure~\ref{fastchem_c_o_z} shows that the $\rm H_2O$ abundance is consistently higher than OH abundance in an equilibrium chemistry model. 
The retrieval results from Sec~\ref{sec:results} indicate that LTT-9779~b's atmosphere cannot be explained only by equilibrium chemistry processes. 
It may experience some disequilibrium processes, such as photochemistry or vertical mixing. 

\subsection{Constraints on OH}

Many studies using HST/WFC3 spectroscopic data have found evidence for hydrides and oxides in the planetary atmosphere \citep{2020AJ....160..112P, 2020AJ....160..109S, 2016ApJ...822L...4E}.
The planets within which OH was detected include terrestrial planets within our solar system, such as Earth \citep{1950ApJ...111..555M}, Venus \citep{2008A&A...483L..29P} and Mars \citep{2013Icar..226..272T}, as well as some exoplanets, such as WASP-33~b \citep{2021ApJ...910L...9N, 2022A&A...668A..53C, 2023AJ....166...31F}, WASP-127~b \citep{2023MNRAS.522.5062B}, WASP-76~b \citep{2021A&A...656A.119L} and WASP-18~b \citep{2023AJ....165...91B} etc. 

Normally, OH in ultra-hot Jupiters is considered to be produced mainly by the thermal dissociation process of $\rm H_2O$, which is expected to occur when day-side temperature exceeds 2000~K. 
Similarly to the study of WASP-33~b by \citet{2021ApJ...910L...9N}, we do not detect the presence of $\rm H_2O$ in LTT-9779~b's atmosphere, indicating that $\rm H_2O$ may have been partly thermally dissociated in its upper atmosphere.
The day-side temperature of LTT-9779~b is higher than 2300~K, based on the 3.6~$\upmu$m secondary eclipse measurement \citep{2020ApJ...903L...6D}. 
\citet{2023A&A...675A..81H} build global climate models to explain the large eclipse depths seen in the CHEOPS observation, and find that the temperature in the lower atmosphere can reach 2500-3000~K. 
Thus, OH could form on the day side of the atmosphere, and transfer to the terminator region by atmospheric circulation, which also helps to send the clouds forming on the night side to the day side \citep{2024ApJ...962L..20R}. Additionally, OH may be detectable through transmission observation due to vertical mixing, which brings it up from deeper and hotter atmosphere layers.

Our Bayesian atmospheric retrievals find no conclusive evidence supporting the presence of OH in LTT-9779 b's atmosphere, and give an upper limit (at 95\% confidence level) of $\log_{10}(\mathrm{OH}) \textless -1.14$ on its VMR.
To confirm or rule out the presence of OH in the atmosphere of this planet, further observations are needed. 
These should include additional transit transmission spectra and emission spectra from secondary eclipses to acquire a more precise T-P profile of the day side and a detailed atmospheric composition of this planet.
High-resolution ground-based infrared spectroscopic observation can also help detect OH in LTT-9779~b.

\begin{figure*}
\includegraphics[width=\textwidth]{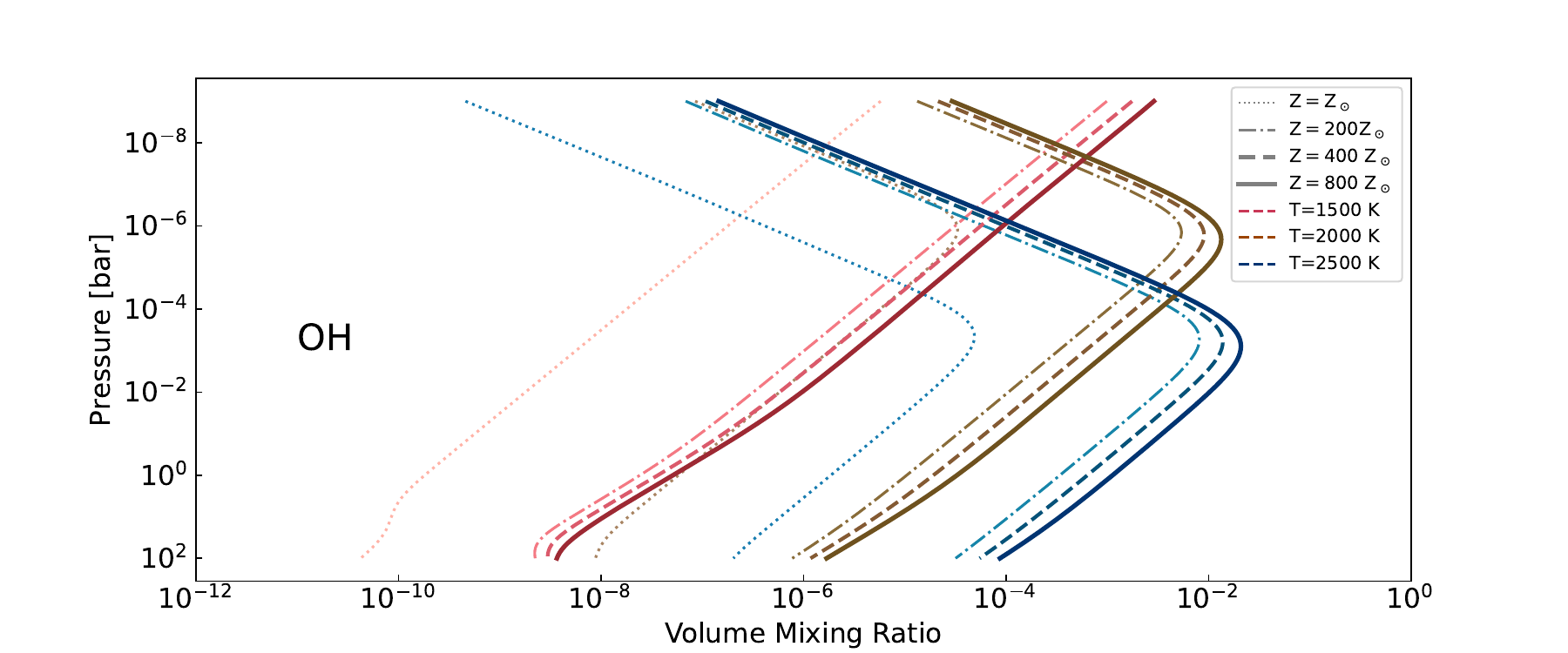}
\caption{The variation of OH VMR calculated assuming equilibrium chemistry with different temperatures and metallicities. 
The different colors represent different temperatures, with red representing $\rm T=1500~K$, brown representing $\rm T=2000~K$, and blue representing $\rm T=2500~K$, respectivley.
The thickness of the lines represents different metallicity, where the thinnest line is used for $\rm Z=Z_\odot$ and the thickest line for $\rm Z=800~Z_\odot$.} 
\label{oh_h2o_p}
\end{figure*}

We also explore the VMR of OH for a given temperature and metallicity at different pressure levels with the assumption of equilibrium chemistry.
According to previous studies of LTT-9779~b (\citealt{2020ApJ...903L...6D, 2023A&A...675A..81H, 2024ApJ...962L..20R, 2025A&A...695A..26R} etc.), it may have a super-solar metallicity and a day-side temperature as high as 2300~K. 
Therefore, to check the VMR distributions under various atmospheric conditions, we set the temperature to 1500~K, 2000~K, and 2500~K, and the metallicity to 1, 200, 400 and 800 times of solar metallicity \citep{2009ARA&A..47..481A}, respectively.  
The pressure is set to vary from $\rm 10^{-9} $ to $\rm 10^2$~bar.
We show the VMR distributions of OH in Figure~\ref{oh_h2o_p}.
From Figure~\ref{oh_h2o_p}, we can see that it is possible to have a high OH abundance with a high atmospheric temperature and super-solar metallicity, and none of the simulated abundance values exceed the OH upper limit determined in this work.

\subsection{Rejection of Temporal Variation Hypothesis}
Previously, \citet{Wilson21} and \citet{Ouyang23} have found possible evidence for temporal variation in the atmospheric compositions of an ultra-hot Jupiter WASP-121~b, which is also predicted by theoretical models \citep{Komacek20}. 
They find that transmission spectra result from different observation runs differ significantly, which can be explained by the change of chemical compositions in the planet terminator region over timescales longer than years. 
Here in our study, we compare the best-fit model spectrum using HST observations obtained in 2021 June and 2022 June with previous observations from Spitzer in 2018 August and September, and also observations from JWST/NIRISS GR700XD in 2022 July to try to find temporal changes in the terminator region of LTT-9779~b. 
\citet{2023A&A...675A..81H} have investigated the possible variability of the eclipse depths of LTT-9779~b, and found no strong evidence due to the low SNR of the CHEOPS observations. 

\begin{figure*}
\includegraphics[width=\textwidth]{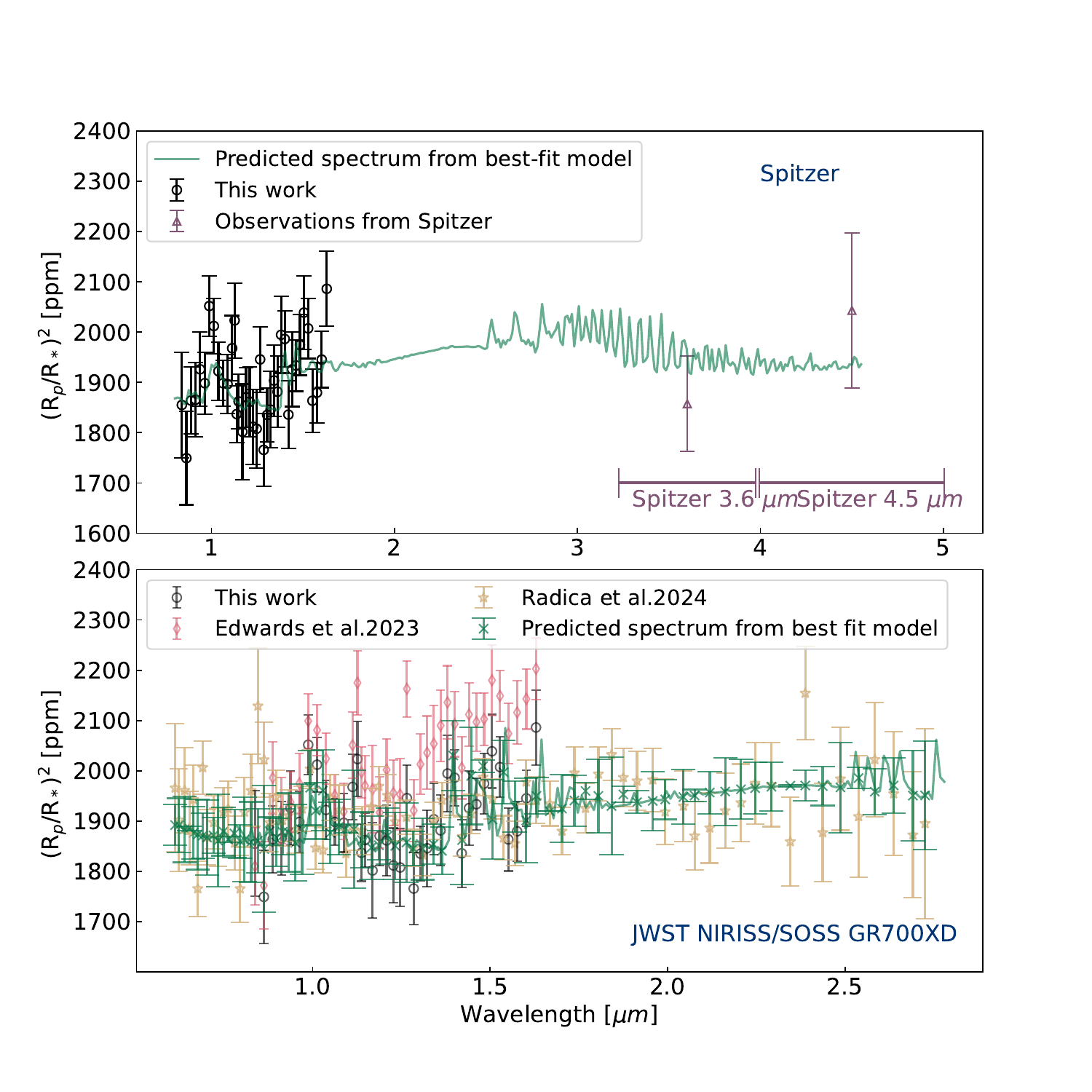}
\caption{Top panel: Theoretic spectrum (green solid line) of LTT-9779 b from the best-fit atmospheric model using HST/WFC3 observations (black data points). 
The purple triangle data points are observations in the Spitzer 3.6~$\upmu$m and 4.5~$\upmu$m bands \citep{2020ApJ...903L...7C}. 
Bottom panel: The green points and line represent the predicted transmission spectrum of LTT-9779~b obtained using one single transit observation from JWST/NIRISS GR700XD, over-plotted with theoretical photon noise. The pink diamonds are observations from \citet{2023AJ....166..158E}. 
The light yellow stars represent observations from \citet{2024ApJ...962L..20R}.}
\label{wfc3-spitzer-jwst}
\end{figure*}

Assuming that the planet has a cloudy H/He dominated primary atmosphere with only OH and FeH as chemical gases, we present the predicted transmission spectrum of LTT-9779~b in a single transit obtained using JWST/NIRISS GR700XD and Spitzer in Figure~\ref{wfc3-spitzer-jwst}. 
In the top panel of Figure~\ref{wfc3-spitzer-jwst}, we show the comparison between the best-fit theoretic spectral model of LTT-9779~b with the observations from Spitzer. 
In the bottom panel, we show the predicted spectrum of JWST/NIRISS GR700XD from the best-fit model and the observations from \citet{2024ApJ...962L..20R}.

From the top panel, we can see that although there is a slight discrepancy between the predicted data point and the actual observation from \citet{2020ApJ...903L...7C} in the Spitzer 3.6~$\upmu$m and 4.5~$\upmu$m band, this discrepancy falls within the uncertainty of the observational data point.
From the bottom panel, it is apparent that the reduced spectrum of HST / WFC3 in this work and the JWST / NIRISS spectrum of \citet{2024ApJ...962L..20R} are almost consistent within the overlapping wavelength range. 
Furthermore, the trend of the predicted spectrum from the best-fit atmospheric model in this work aligns closely with the observations reported by \citet{2024ApJ...962L..20R}, despite minor differences between them. 

The overall consistency between the observations from HST/WFC3 in this work, the predicted spectrum based on the best-fit model from HST/WFC3, and observations from both Spitzer and JWST provides evidence that the planetary atmosphere has not undergone significant temporal variations in the terminator region.

\section{Conclusion}
\label{sec:conclusions}
In this work, we present an analysis of transit spectroscopic observations of LTT-9779~b, an ultra-hot Neptune located in the hot-Neptune Desert, obtained using HST/WFC3. 
We use the state-of-the-art open-source data reduction pipeline \texttt{Iraclis} and the atmospheric retrieval code \texttt{TauREx3} to analyze the data. 
The transmission spectrum is best fitted with a cloudy H/He-dominated primary atmospheric model that includes chemical gases OH and FeH. 
We can reject the pure cloudy flat-line model with $\sim$ 2.7-$\sigma$ confidence.
Since the statistical significance for the presence of OH and FeH is low, we give the 95\% confidence level upper limits on the OH and FeH VRMs at $7.18\times10^{-2}$ and $1.52\times10^{-8}$, respectively.

The difficulty of modeling this planet's atmospheric composition using equilibrium chemistry suggests that disequilibrium mechanisms, such as vertical mixing and photochemistry, may be at play in the terminator region of LTT-9779~b. Additionally, we can reject the hypothesis of significant temporal variations in the atmospheric properties of its terminator region.

LTT-9779~b remains an interesting target for future planetary atmosphere observations, both from ground-based high-resolution and space-based lower-resolution facilities, due to its unresolved mysteries. 
Further ground-based high-resolution observations could test the constraints on OH and FeH obtained by this work.
Additionally, the emission spectrum from the day side will be instrumental in constraining the T-P profile and the chemical composition of the hot hemisphere.
%, potentially explaining the upper limits of OH and FeH. 
Combining observations from multiple facilities, including the future CSST mission, Tianlin mission \citep{2023RAA....23i5028W} and Ariel Mission \citep{2018ExA....46..135T}, which will provide moderate to high-resolution transmission and emission spectra on a population of exoplanets, with a broader wavelength coverage could better quantitatively constrain the atmospheric composition and metallicity of this mysterious planet.

\section*{acknowledgements}
We thank our anonymous referee for prompt and insightful comments which led to significant improvement of the manuscript. 
We thank professor Giovanna Tinetti, whose visit triggered this study. 
We thank Angelos Tsiaras, Ingo Waldmann, and Ahmed Al-Refaie for their instructions on how to use \texttt{Iraclis} and \texttt{TauREx}. 
We acknowledge support from the National Key R\&D Program of China, No. 2024YFA1611800, 
the Postdoctoral Fellowship Program of CPSF (GZB20230767), 
Pre-research project on Civil Aerospace Technologies No. D010301 funded by China National Space Administration (CNSA), the National Natural Science Foundation of China (NSFC) under grant Nos. 12073092, 12103097, 12103098, 12063001, 42075123, 11988101 and 62127901, the science research grants from the China Manned Space Project (No. CMS-CSST-2021-B09, No. CMS-CSST-2021-B12), Provincial Natural Science Foundation of Hainan under grant No. 424QN217, and the ET2.0 mission.
L.Z. is supported by the Chinese Academy of Sciences (CAS), through a grant to the CAS South America Center for Astronomy (CASSACA) in Santiago, Chile.
This research was also supported by the China Scholarship Council (CSC). I would like to express my sincere gratitude to the CSC for providing the financial support that enabled me to pursue my studies and complete this research.

This work is based on observations with the NASA/ESA Hubble Space Telescope,  obtained at the Space Telescope Science Institute (STScI) operated by AURA, Inc. 
The publicly available HST observations presented here were taken as part of proposal 16457, led by Billy Edwards. 
These were obtained from the Hubble Archive, which is part of the Mikulski Archive for Space Telescopes. 

\section*{Data availability}
The data underlying this article will be shared on reasonable request to the corresponding author.

\bibliography{ltt-9779b}
\bibliographystyle{aasjournal}

\appendix

\section{additional tables and figures}

\setcounter{figure}{0}
\renewcommand{\thefigure}{A\arabic{figure}}

\begin{figure*}
\includegraphics[width=0.48\textwidth]{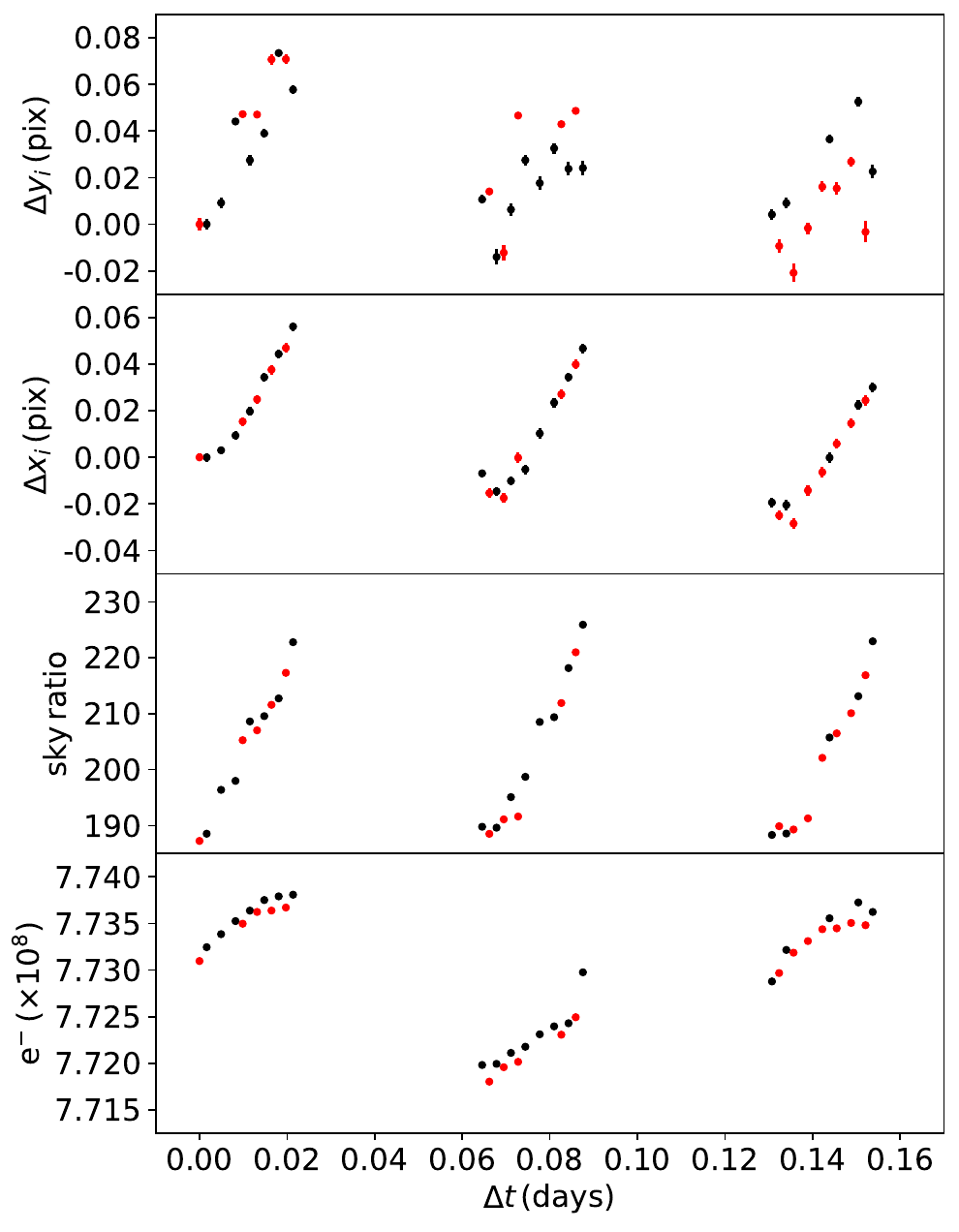}{(a)}
\includegraphics[width=0.48\textwidth]{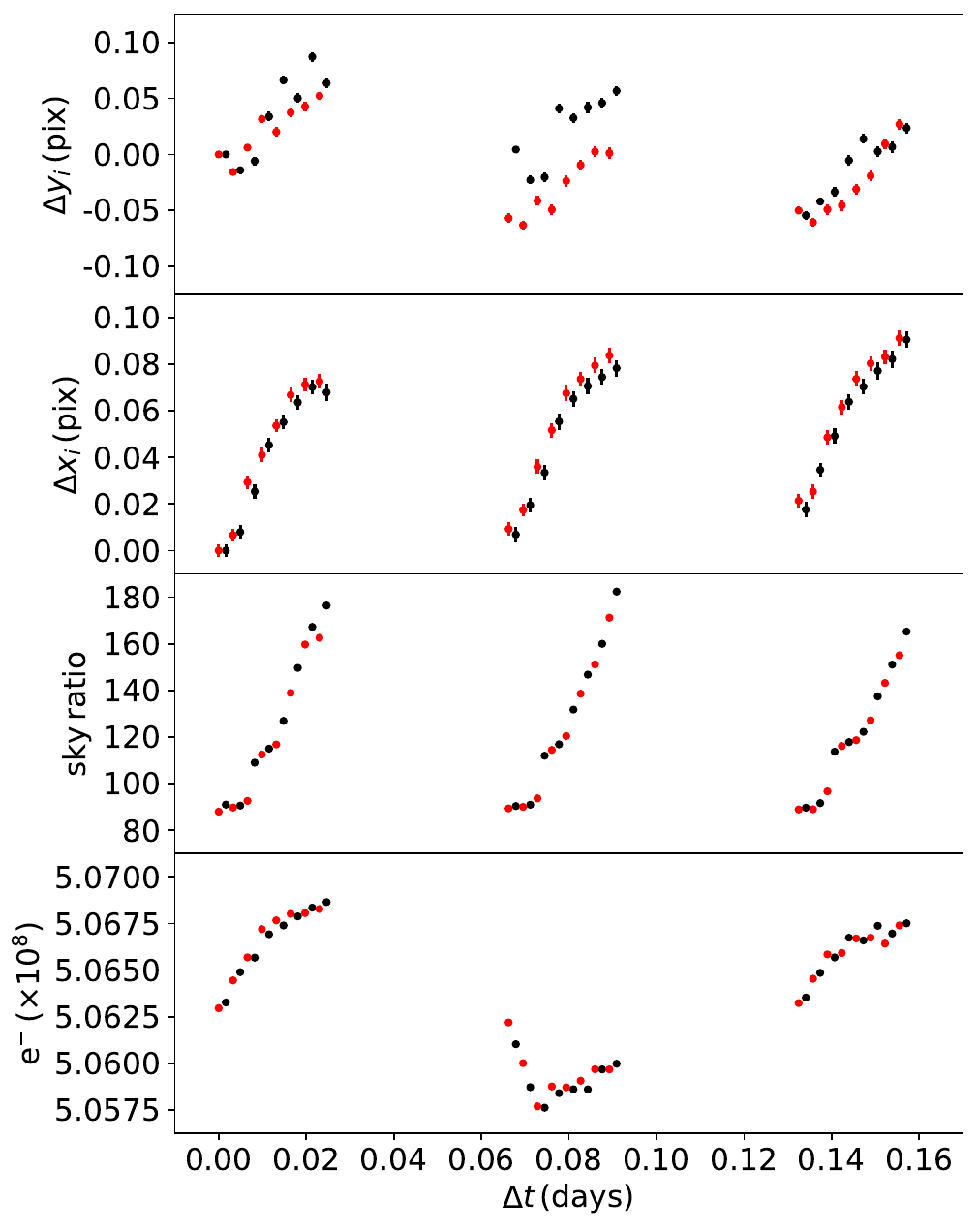}{(b)}
\caption{Vertical ($\Delta y$) and horizontal ($\Delta x$) shifts for each spectral image relative to the first one, the sky ratio and raw white flux of each image in the G141 grism visit (panel a) and the G102 grism visit (panel b). Black dots show the forward scan and red dots show the reverse scan. }
\label{diagnostics}
\end{figure*}

\begin{figure*}[p]
\includegraphics[width=0.6\textwidth]{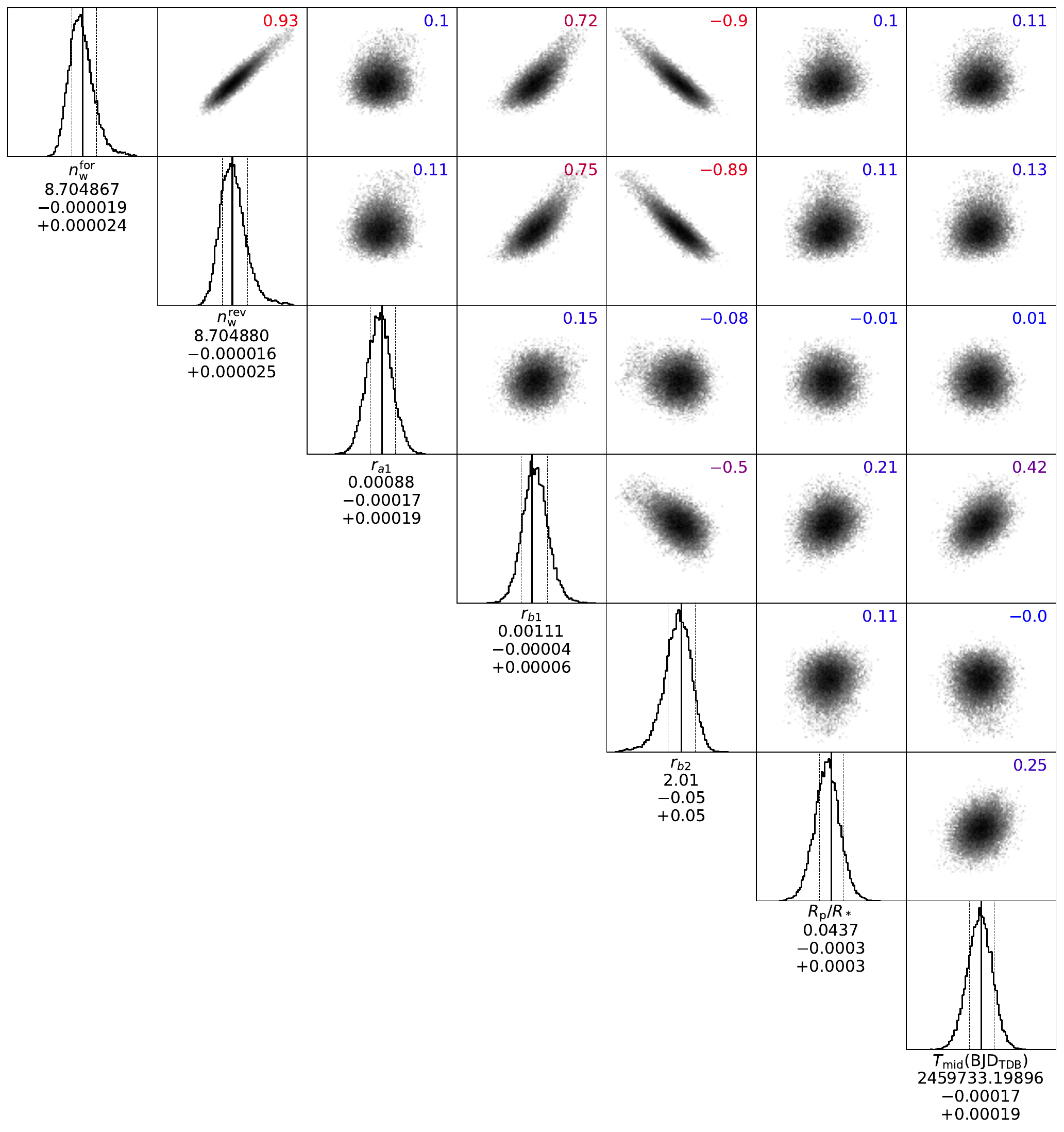}{(a)}
\includegraphics[width=0.6\textwidth]{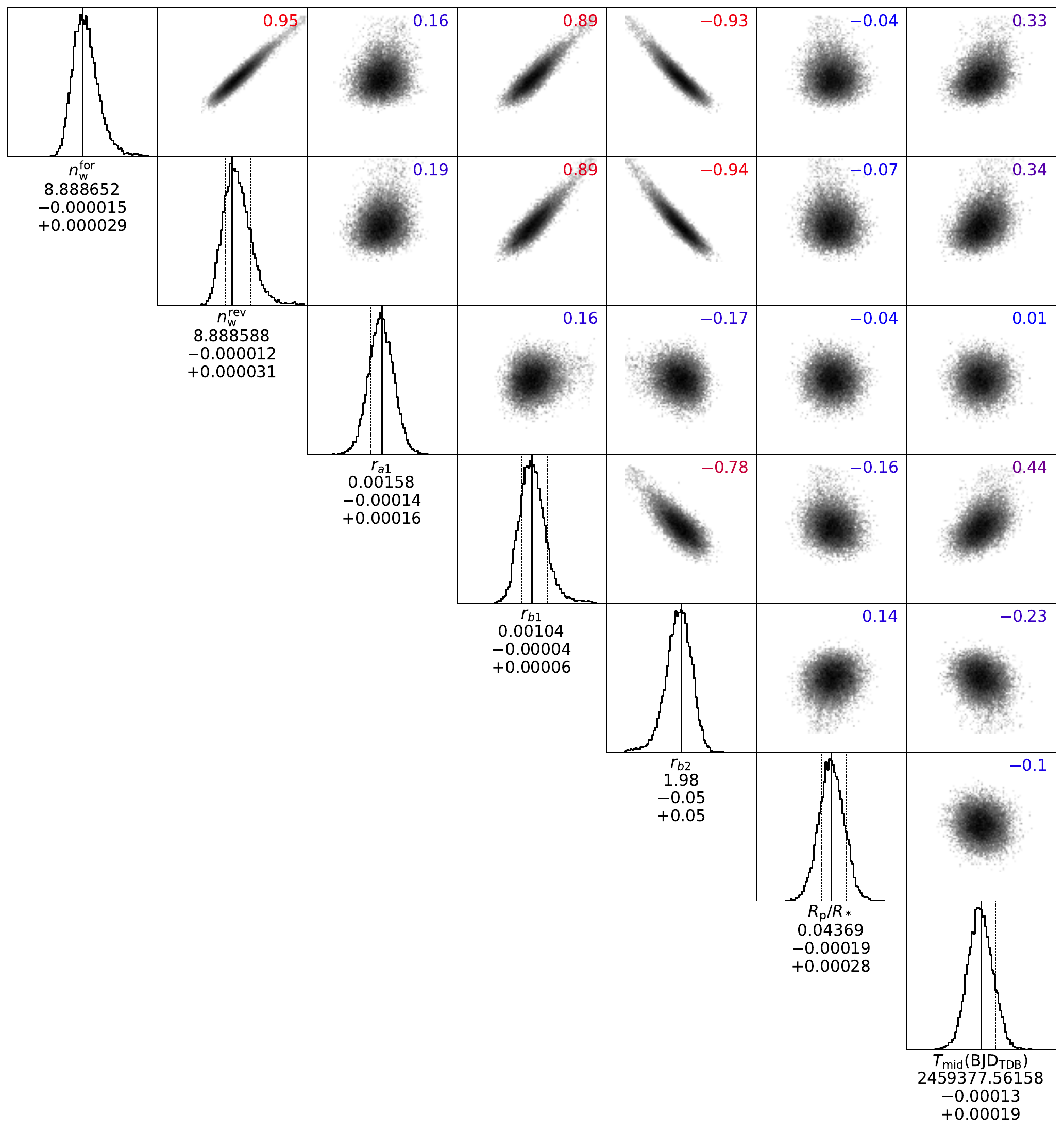}{(b)}
\caption{Corner plot showing the correlations between the fitted systematics and transit parameters for the HST/WFC3 G102 (panel a) and G141 (panel b) data. 
The plots demonstrate that the radius ratio, $\rm R_p/R_*$, is not strongly correlated with any of the systematic parameters.}
\label{white_correlations}
\end{figure*}

\begin{figure*}
\includegraphics[width=1.0\textwidth]{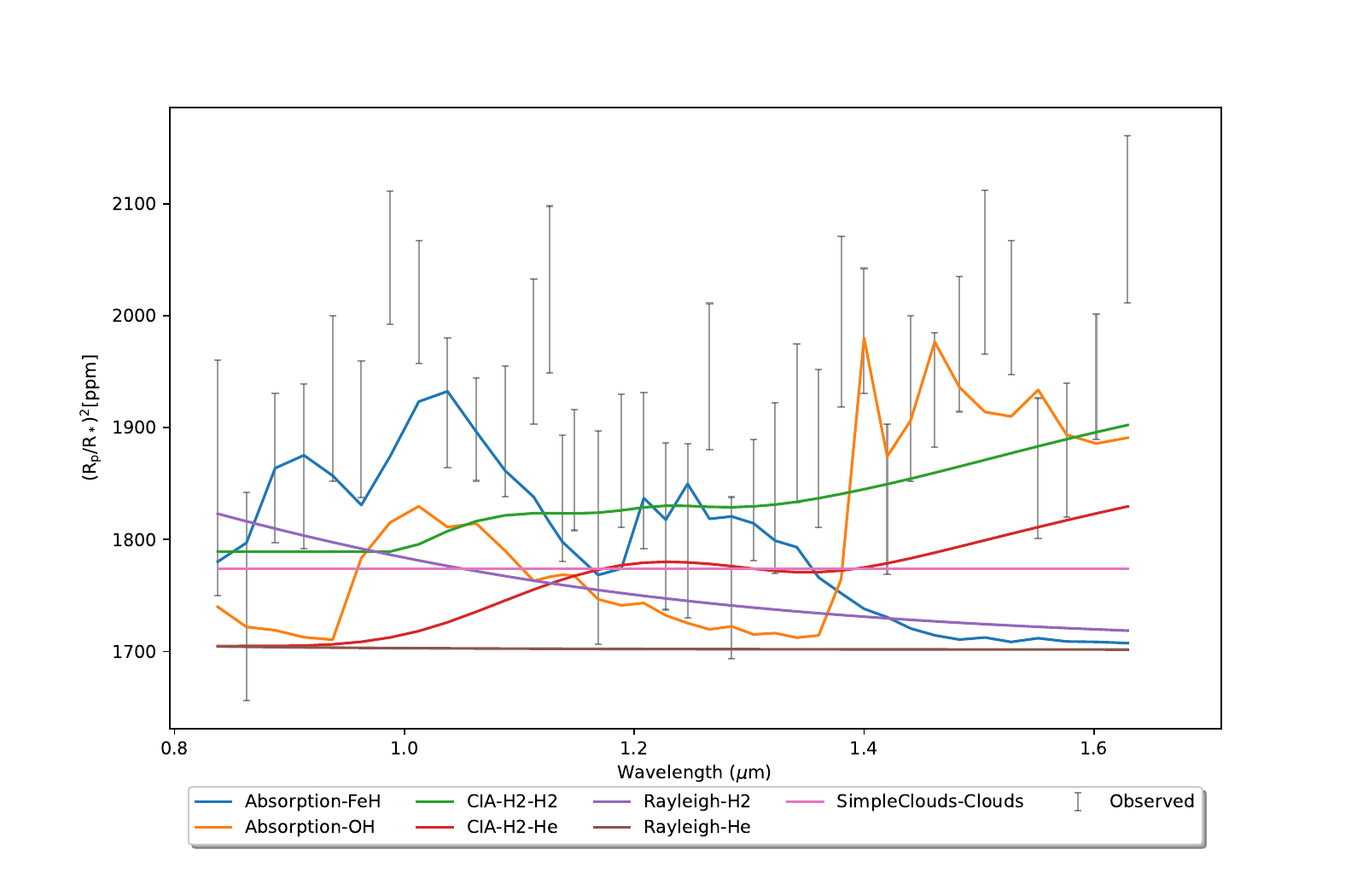}
\caption{
Decomposition of atmospheric contributions in the \texttt{TauREx} retrieval framework for LTT-9779~b, which includes clouds, CIA, Rayleigh scattering, and chemical gases. The plot highlights the individual contributions of these components on the observed spectrum.}
\label{contributions_hst_full_model}
\end{figure*}

\begin{figure*}
\includegraphics[width=1.0\textwidth]{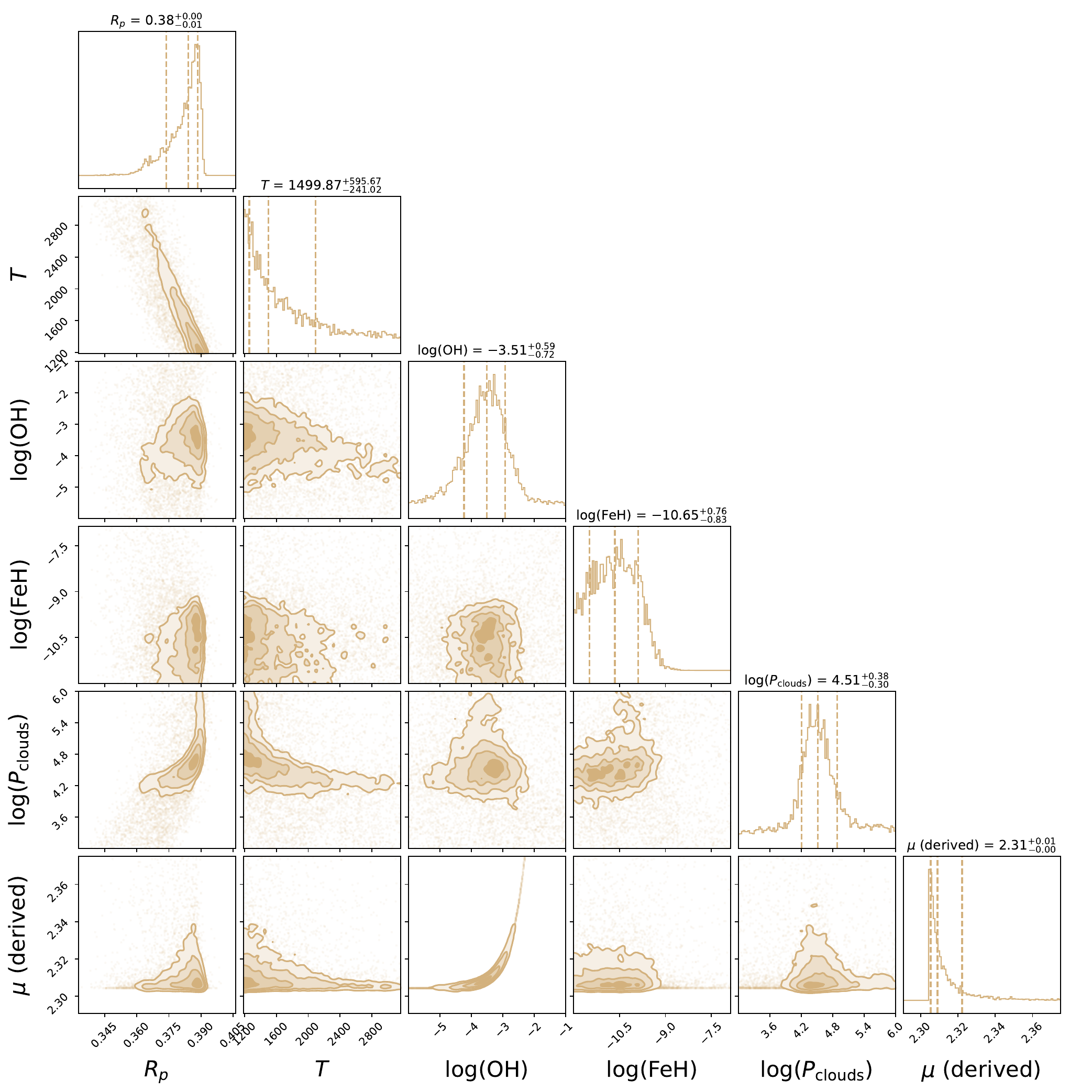}
\caption{Similar to Figure~\ref{posterior_g102_g141}, but for combined observations from HST/WFC3 and JWST data from \citet{2024ApJ...962L..20R}.}
\label{posterior_hst_jwst}
\end{figure*}

% \clearpage
% \section{additional tables}
\setcounter{table}{0}
\renewcommand{\thetable}{A\arabic{table}}
\begin{table*}
\centering
\small
\caption{Statistical values of different free chemistry models, including  Bayesian log evidences (ln(E)), significance levels (Sigma) compared to the flat-line model, and logarithmic Bayes factor ($\Delta$ln(E)).}
\label{retrieval_results_constant}
\begin{threeparttable}
\resizebox{0.68\textwidth}{!}{
\begin{tabular}[b]{llll}
\hline
models & $\rm ln$(E) & Sigma& $\Delta \rm ln(E)$\\
\hline
Flat-line &300.488&-&-\\
Clouds+Rayleigh+CIA&300.904 &1.599&0.416\\
\hline
Single molecule & ln(E) & Sigma&$\Delta \rm ln(E)$\\
&&(compared with flat-line)& (compared with flat-line model)\\
OH&302.301&2.455&1.813\\
CP&301.574&2.075&1.086\\
FeH&301.426&1.985&0.938\\
$\rm CO_2$&301.397&1.967&0.909\\
$\rm C_2H_4$&301.180&1.820&0.692\\
$\rm H_2O$&301.034&1.709&0.546\\
AlH&300.985&1.669&0.497\\
CO&300.948&1.637&0.460\\
$\rm CH_4$&300.925&1.617&0.437\\
$\rm NH_3$&300.817&1.516&0.329\\
MgH&300.787&1.485&0.299\\
HCN&300.782&1.480&0.294\\
$\rm C_2H_2$&300.707&1.396&0.219\\
$\rm H_2CO$&300.695&1.381&0.207\\
AlO&300.653&1.327&0.165\\
CaH&300.543&1.140&0.055\\
TiH&300.497&0.995&0.009\\
ScH&300.386&-&-0.102\\
CN&300.342&-&-0.146\\
MgO&300.268&-&-0.220\\
TiO&300.253&-&-0.235\\
CrH&300.218&-&-0.270\\
VO&300.050&-&-0.438\\
\hline
Two molecules&$\rm ln$(E) & Sigma&$\Delta \rm ln(E)$\\
(OH+)&&(compared with flat-line)& (compared with OH-only model)\\
CP&302.693&2.631&0.392\\
FeH&302.883&2.711&0.582\\
$\rm CO_2$&302.417&2.509&0.116\\
$\rm C_2H_4$&302.737&2.650&0.436\\
$\rm H_2O$&302.021&2.319&-0.280\\
AlH&302.395&2.499&0.094\\
CO&302.239&2.426&-0.062\\
$\rm CH_4$&302.118&2.367&-0.183\\
$\rm NH_3$&302.111&2.364&-0.190\\
MgH&302.520&2.555&0.219\\
HCN&301.992&2.304&-0.309\\
$\rm C_2H_2$&301.983&2.299&-0.318\\
$\rm H_2CO$&302.114&2.365&-0.187\\
AlO&302.059&2.338&-0.242\\
CaH&301.805&2.206&-0.496\\
TiH&302.151&2.383&-0.150\\
ScH&301.687&2.140&-0.614\\
CN&302.013&2.315&-0.288\\
MgO&301.440&1.994&-0.861\\
TiO&301.700&2.148&-0.601\\
CrH&301.662&2.126&-0.639\\
VO&301.419&1.981&-0.882\\
\hline
final model&$\rm ln(E)$ &Sigma&$\Delta \rm ln(E)$\\
&&(compared with flat-line)& (compared with flat-line)\\
OH+FeH&302.883&2.711&2.395\\
\hline
\end{tabular}}
\end{threeparttable}
\end{table*}
\label{lastpage}
\end{document}